\documentclass[12pt]{article}
\usepackage{epsf,epsfig,amssymb,amsmath,graphicx,float,latexsym} 

\newcommand{\lsim}{\raisebox{-0.13cm}{~\shortstack{$<$ \\[-0.07cm] $\sim$}}~}
\newcommand{\gsim}{\raisebox{-0.13cm}{~\shortstack{$>$ \\[-0.07cm] $\sim$}}~}

\begin{document}
\renewcommand{\thefootnote}{\fnsymbol{footnote}}

\begin{titlepage}
  
\begin{flushright}
October 2008
\end{flushright}

\begin{center}

\vspace{1cm}

{\Large {\bf Neutralino Dark Matter in an SO(10) Model with Two-step
    Intermediate Scale Symmetry Breaking}}

\vspace{1cm}

{\sc Manuel Drees}\footnote{drees@th.physik.uni-bonn.de} and
{\sc Ju Min Kim}\footnote{juminkim@th.physik.uni-bonn.de} \\

\vskip 0.15in
{\it
{Physikalisches Institut der Universit\"at Bonn,
Nussallee 12, 53115 Bonn, Germany}\\
}
\vskip 0.5in

\abstract{We consider a supersymmetric Grand Unified Theory (GUT) based on the
  gauge group $SO(10)$ suggested by Aulakh et al., which features two--step
  intermediate symmetry breaking, $SO(10) \rightarrow SU(4)_C \times SU(2)_L
  \times SU(2)_R \rightarrow SU(3)_C \times U(1)_{B-L} \times SU(2)_L \times
  SU(2)_R \rightarrow SU(3)_C \times SU(2)_L \times U(1)_Y$. {\bf $45, 54,
    126+\overline{126}$} dimensional representations of Higgs superfields are
  employed to achieve this symmetry breaking chain. We also introduce a
  second, very heavy, pair of Higgs doublets, which modifies the Yukawa
  couplings of matter fields relative to minimal $SO(10)$ predictions. We
  analyze the differences in the low energy phenomenology compared to that of
  mSUGRA, assuming universal soft breaking scalar masses, gaugino masses and
  trilinear couplings at the GUT scale. We find that thermal neutralino Dark
  Matter remains viable in this scenario, although for small and moderate
  values of $\tan\beta$ the allowed region is even more highly constrained
  than in mSUGRA, and depends strongly on the the light neutrino masses.}

\end{center}
\end{titlepage}
\setcounter{footnote}{0}

\section{Introduction}

Grand Unified Theories (GUTs) based on the gauge group $SO(10)$
\cite{so10old,gn2} have been investigated extensively. This choice of
gauge group has several appealing features. First of all, it has room for a
right-handed neutrino per generation in the 16--dimensional irreducible spinor
representation which includes all known matter fields. Thus it provides a
beautiful explanation of the smallness of the neutrino mass via the ``seesaw
mechanism'' \cite{minkowski}.  Moreover, the existence of very massive
right--handed neutrinos might also allow to explain the asymmetry between
matter and antimatter in the Universe by thermal leptogenesis \cite{fy2}.
Furthermore, $SO(10)$ contains the ``Pati--Salam'' \cite{ps} group $SU(4)_C
\times SU(2)_L \times SU(2)_R$ as subgroup, meaning that parity is preserved
at high energy and broken spontaneously.

On the other hand, the fact that the rank of $SO(10)$ is five causes some
complications. Recall that the rank of the Standard Model (SM) gauge group
$G_{\rm SM} = SU(3)_C \times SU(2)_L \times U(1)_Y$ is only four. There are
several ways of breaking $SO(10)$ down to $G_{\rm SM}$, depending on which
representations of Higgs fields are introduced in the theory. Here we consider
the possibility of having intermediate phase(s) at energy scales well below
the GUT scale. The existence of a scale near $10^{14}$ GeV can be motivated by
neutrino oscillation experiments \cite{sno,superk,kamland}: the mass of the
heaviest neutrino cannot be less than $\sqrt{\delta m_{atm}^2} \sim 0.04$ eV.
In the seesaw mechanism this translates into an upper bound on the
right--handed Majorana neutrinos mass if we assume that the largest neutrino
Yukawa coupling is order unity, $M_N \lesssim 10^{14}$ GeV. Note that $M_N$
breaks the $SU(2)_R$ subgroup of $SO(10)$. It thus seems natural to assume the
left--right symmetric subgroup of $SO(10)$ to be broken to $G_{\rm SM}$ near
this scale (``$M_R$''), if we assume that the Yukawa coupling that gives rise
to the Majorana mass $M_N$ is also of order unity.

In this work, we will analyze the consequences of this assumption, by
considering the low energy phenomenology of a supersymmetric $SO(10)$ model
suggested by Aulakh et al. \cite{aulakh}. It features the symmetry breaking
chain
\begin{equation} \label{chain}
SO(10) \xrightarrow[M_X]{54} G_{422D} \xrightarrow[M_C]{45}
  G_{3122} \xrightarrow[M_R]{{126}+\overline{126}} G_{\rm SM}\, .
\end{equation}
Here we have used the notation $G_{3122} = SU(3)_C \times U(1)_{B-L} \times
SU(2)_R \times SU(2)_L$ and $G_{422D} = SU(4)_C \times SU(2)_R \times SU(2)_L
\times D$, where $D$ is a discrete symmetry which ensures that $SU(2)_L$ and
$SU(2)_R$ have equal gauge couplings. We assume universal (``mSUGRA''
\cite{msugra}) boundary conditions for the soft supersymmetry breaking terms
at the GUT scale $M_X$. This means that all soft breaking scalar masses are
equal to $m_0$ at the GUT scale, while all gaugino masses are equal to
$M_{1/2}$; moreover, all SUSY breaking trilinear scalar couplings are
characterized by the single parameter $A_0$.

Introducing two intermediate scales, and the corresponding additional gauge,
matter and Higgs superfields, has three main effects. First, the right--handed
neutrinos obtain Majorana masses at scale $M_R$ by coupling to the {\bf
  126}--dimensional Higgs whose vacuum expectation value (VEV) is responsible
for breaking $G_{3122}$ in Eq.(\ref{chain}). These Majorana Yukawa couplings,
as well as the extra Dirac couplings of the light neutrinos, will change the
low energy spectrum of soft breaking parameters via renormalization group
equations (RGEs). Secondly, since we have to introduce many more additional
Higgs than gauge superfields to achieve the symmetry breaking chain
(\ref{chain}), all gauge couplings increase quite rapidly at high energy
scales $\gsim M_R$. As a result the gaugino masses, which we assume to be
universal at $M_X$, decrease significantly when they evolve down to $M_R$.
Finally, the enhanced gauge symmetry at energies $\gsim M_R$ also increases
the size of gauge contributions to the RGE of all scalar masses. Note that the
second and third effect tend to cancel, if the scalar masses are expressed in
terms of the GUT--scale input parameter $m_0$ and $M_{1/2}$.

The rest of this paper is organized as follows. In the next Section, we review
the main features of the model \cite{aulakh} we are considering. We also
describe the numerical methods used in our analysis. In Sec.~3 we discuss the
most important experimental and cosmological constraints on the parameter
space of the model. Our numerical results are given in Sec.~4. Special
attention is devoted to the regions of parameter space where the lightest
neutralino makes a good thermal Dark Matter candidate in standard cosmology.
Finally, we conclude in Sec.~5.

\section{The Set-Up}

\subsection{The model}

We will consider the model suggested by Aulakh et al. \cite{aulakh}. It is
based on the gauge group $SO(10)$. Besides three generations of matter
superfields residing in {\bf 16}--dimensional representations as well as the
{\bf 45}--dimensional gauge superfields, we introduce Higgs superfields in the
{\bf 54, 45, 126}, $\overline{{\bf 126}}$ and {\bf 10} representations of
$SO(10)$. The Higgs superfields required to break $SO(10)$ down to $G_{\rm
  SM}$ can be described by the tensors
\begin{eqnarray} \label{tensors}
{\bf 54} &:& S_{ij} = S_{ji} \ {\rm and} \ S_{ii} = 0\,,\quad {\bf 45}: A_{ij}
= - A_{ji}\,,  
\nonumber \\ 
{\bf 126} &:& \Sigma_{ijklm} = \frac{i}{5!} \epsilon_{ijklmopqrs}
\Sigma_{opqrs}\,, 
\nonumber \\
\overline{{\bf 126}} &:& \overline{\Sigma}_{ijklm} =
-\frac{i}{5!} \epsilon_{ijklmopqrs} \overline{\Sigma}_{opqrs}\, 
\end{eqnarray}
where the subscripts $i,j,k,\dots$ run from 1 to 10, and repeated subscripts
are summed.

This allows us to realize the symmetry breaking chain (\ref{chain}) with a
purely renormalizable superpotential, given by \cite{aulakh}
\begin{equation} \label{WSSB}
W_{SSB} = \frac{m_S}{2}\mathrm{tr} S^2 + \frac{\lambda_S}{3} \mathrm{tr} S^3 
+ \frac{m_A}{2} \mathrm{tr} A^2 + \lambda \mathrm{tr} A^2 S 
+ m_\Sigma \Sigma \overline{\Sigma} + \eta_S \Sigma^2 S +
\overline{\eta}_S \overline{\Sigma}^2 S + \eta_A \Sigma \overline{\Sigma} A\,
. 
\end{equation}
A crucial observation \cite{fittest,aulakh} is that some components of the
Higgs superfields listed in (\ref{tensors}) are much lighter than one might
naively expect. For example, even though the {\bf 45}--plet $A$ is responsible
for the breaking of $G_{422D}$ to $G_{3122}$ at scale $M_C$, some components
of $A$ only acquire masses of order $M_C^2 / M_X$ or $M_R^2 / M_C$, whichever
is larger. Similarly, even though $\Sigma$ and $\overline{\Sigma}$ are
responsible for breaking $G_{3122}$ to the SM gauge group, some of their
components only get masses of order $M_R^2/M_X$. On the other hand, some
components of $A, \Sigma$ and $\overline{\Sigma}$ obtain masses of order
$M_X$.

This is summarized in Table \ref{table:mass}. Here we have used the
decompositions of the Higgs fields under $SU(4)_C \times SU(2)_L \times
SU(2)_R$:
\begin{eqnarray} \label{decom}
S &=& (1,1,1) \oplus (20,1,1) \oplus (1,3,3) \oplus (6,2,2) \, ;
\nonumber \\
A &=& (15,1,1) \oplus (1, 1, 3) \oplus (1,3,1) \oplus (6,2,2) \, ;
\nonumber \\
\overline{\Sigma} &=& (10,1,3) \oplus (\overline{10}, 3, 1) \oplus (15,2,2) 
\oplus (6,1,1)\, ; 
\nonumber \\
\Sigma &=& (\overline{10}, 1, 3) \oplus (10, 3, 1) \oplus (15,2,2)
\oplus (6, 1, 1) \,.
\end{eqnarray}
The components of the Higgs fields that acquire large vacuum expectation
values (vevs) appear as the first term in each right--hand side (rhs) of
Eqs.(\ref{decom}); in addition, the $(1,1,3)$ component of $A$ is also assumed
to obtain a nonzero vev \cite{aulakh}.

\begin{table}[h]
\begin{center}
\begin{tabular}{|l|l|}
\hline
\hspace{2cm}State & \hspace{1cm} Mass \\
\hline 
all of $S$ & \\
all of $A$, except $(15,1,1)_A$ & $\sim M_X$ \\
all of $\Sigma$ and $\overline\Sigma$, except $SU(4)_C$ (anti--)decuplets
&   \\
\hline
$(\overline{10},3,1)_{\overline \Sigma}$ and $({10},3,1)_{\Sigma}$ &  \\
color triplets and sextets of $(10,1,3)_{\overline \Sigma}$  and
$(\overline{10},1,3)_{\Sigma}$ & $\sim M_C$ \\
color triplets of $(15,1,1)_A$ & \\
\hline
$(\delta^0-\overline\delta^0),\quad \delta^+,\quad \overline\delta^- $
& $\sim M_R$ \\ \hline
color octet and singlet of $(15,1,1)_A$
&  $\sim M_1 \equiv {\rm max} \left[ {M_R^2\over M_C}, {M_C^2\over M_X}
\right]$  
\\ \hline
$(\delta^0+\overline\delta^0),\quad \delta^{++},\quad
\overline\delta^{--}$ & $\sim M_2 \equiv {M_R^2 / M_X}$ \\
\hline
\end{tabular}
\end{center}
\caption{The spectrum of Higgs superfields after symmetry breaking. The Higgs
 superfields have been introduced in Eq.(\ref{tensors}), and their
 decomposition into irreducible representations of $SU(4)_C \times SU(2)_L
 \times SU(2)_R$ is given in Eq.(\ref{decom}). $\delta^{0,+,++}$ form the
 color singlet part of the $(\overline{10},1,3)$ component of $\Sigma$, while 
$\bar \delta^{0,-,--}$ form the color singlet part of $(10,1,3)$ of
 $\overline{\Sigma}$. Adapted from ref.\cite{aulakh}.}
\label{table:mass}
\end{table}

We also need Higgs superfields in the {\bf 10}--dimensional representation of
$SO(10)$ to provide the Higgs doublet superfields of the Minimal
Supersymmetric Standard Model (MSSM) that break the electroweak gauge symmetry.
Minimal $SO(10)$, with a single {\bf 10}, would require all Yukawa couplings
of one generation to unify, which leads to wrong predictions for ratios of
quark and lepton masses.\footnote{This prediction can be made to work for the
  third generation, if the ratio of MSSM Higgs vevs $\tan\beta$ is large and
  sfermion masses lie well above a TeV \cite {yukuni}; however, they will fail
  for the first two generations.} Introducing $G_{422D}$ as symmetry group
between $M_X$ and $M_C$ aggravates this problem, since it predicts Yukawa
unification at scale $M_C$ if both MSSM Higgs doublets reside in a single {\bf
  (1,2,2)} of $SU(4)_C \times SU(2)_L \times SU(2)_R$.  We therefore include
two such superfields. We assume that the additional bidoublet obtains a mass
through the coupling to the $(1,3,1)$ of $A$, in which case its mass will be
of order $M_2 = M_R^2/M_X$ \cite{aulakh}.

Let us discuss the structure of the matter Yukawa couplings in a bit more
detail. Here we are only interested in third generation couplings, which can
be large enough to affect the weak--scale sparticle spectrum significantly.
The Yukawa unification conditions we will derive will not work for first and
second generation fermions. We assume that this problem is solved by
introducing some more complicated flavor structures, e.g. via
non--renormalizable terms, without introducing additional large couplings.

At energy scales below $M_2$ we have the well--known MSSM superpotential,
\begin{equation} \label{Y_MSSM}
W_{\rm Yuk, MSSM} = Y_u U^c Q H_u + Y_d D^c Q H_d + Y_e E^c L H_d\,,
\end{equation}
where $Q$ and $L$ are the quark and lepton doublets, $U^c,\,D^c$ and $E^c$ the
corresponding singlets, and $H_u$ and $H_d$ the two Higgs doublet
superfields. In Eq.(\ref{Y_MSSM}) we have suppressed all generation and group
indices. 

At energies above $M_2$ the second pair of Higgs doublets as well as some
parts of the $SU(2)$ triplet Higgs superfields (see Table 1) become
accessible. A general ansatz for the matter superpotential is then
\begin{equation} \label{Y_gen}
W_{\rm Yuk, gen} = \sum_{i=1}^2 \left( Y_{u,i} U^c Q H_{u,i} + Y_{d,i} D^c Q
  H_{d,i} + Y_{e,i} E^c L H_{d,i} \right) + \frac {1}{2} Y_N E^c \bar
  \delta^{--} E^c \,.
\end{equation}
The last term in Eq.(\ref{Y_gen}) results from the interaction giving rise to
large Majorana masses for the right--handed neutrino superfields (see below).
The light Higgs doublets $H_u$, $H_d$ are mixtures of the Higgs superfields
appearing in Eq.(\ref{Y_gen}):
\begin{eqnarray} \label{mix}
H_u &=& \cos \varphi_u H_{u,1} + \sin \varphi_u H_{u,2}\,; \nonumber \\
H_d &=& \cos \varphi_d H_{d,1} + \sin \varphi_d H_{d,2}\,.
\end{eqnarray}
At scales above $M_R$, $U^c$ and $D^c$ form a doublet $Q^c$ of $SU(2)_R$;
similarly, the right--handed neutrino superfield $N^c$ and $E^c$ form an
$SU(2)_R$ doublet $L^c$.\footnote{Note that $Q^c$ and $L^c$ are independent
  left--chiral superfields, {\em not} the charge conjugates of $Q$ and $L$.}
Moreover, the Higgs superfields $H_{u,i}, \, H_{d,i}$ are grouped into two
bidoublets $\Phi_i$. Finally, at this scale all members of the $SU(2)_R$
triplet Higgs superfield $\bar\delta$ become accessible. The superpotential
(\ref{Y_gen}) then becomes
\begin{equation} \label{Y_3122}
W_{\rm Yuk, 3122} = \sum_{i=1}^2 \left( Y_{q,i} Q^c Q \Phi_i + Y_{l,i} L^c L
  \Phi_i \right) + \frac{1}{2} Y_N L^c \bar{\delta} L^c\,.
\end{equation}
The last term in Eq.(\ref{Y_3122}) gives rise to large Majorana masses for the
$N^c$ once the neutral component of the $SU(2)_R$ triplet $\bar{\delta} \in
\overline{\Sigma}$ gets a vev. Finally, at scales above $M_C$, $Q$ and $L$
are unified into $F$ in the {\bf(4,2,1)} representation of $G_{422}$, while
$Q^c$ and $L^c$ join to form $F^c$ in the {($\overline{\bf 4}, {\bf 1, 2}$)}
representation. One is then left with a single Yukawa coupling per Higgs
bidoublet,
\begin{equation} \label{Y_422}
W_{\rm Yuk, 422} = \sum_{i=1}^2 Y_i F^c F \Phi_i + \frac{1}{2} Y_N \left(
F^c \overline{\Sigma}_R F^c + F \overline{\Sigma}_L F \right) \,,
\end{equation}
where $\overline{\Sigma}_R$ and $\overline{\Sigma}_L$ are in the 
{(${\bf 10}, {\bf 1, 3}$)} and {($\overline{\bf 10}, {\bf 3, 1}$)}
representation, respectively, of $SU(4)_C \times SU(2)_L \times SU(2)_R$; the
last term in Eq.(\ref{Y_422}) also has to have coupling $Y_N$ due to the
discrete $D$ symmetry.
  
As a first simplification, let us work in the basis where $Y_2 = 0$. This can
always be accomplished by a unitary rotation between the two $\Phi_i$. Since
superpotential couplings renormalize multiplicatively, this choice is
renormalization scale invariant. It is then easy to see that, through the
matching conditions at $M_C$, $Y_{q,2} = Y_{l,2} = 0$ in Eq.(\ref{Y_3122});
similarly, matching at scale $M_R$ implies $Y_{u,2} = Y_{d,2} = Y_{e,2} = 0$
in Eq.(\ref{Y_gen}). The sums in Eqs.(\ref{Y_gen}) and (\ref{Y_3122}) thus
also collapse to single terms. Inserting Eqs.(\ref{mix}) into Eq.(\ref{Y_gen})
then leads to the following matching conditions for the MSSM Yukawa couplings
at scale $M_2$:
\begin{equation} \label{match}
Y_{u,1} = Y_u / \cos \varphi_u\,; \quad Y_{d,1} = Y_d / \cos \varphi_d\,;
\quad Y_{e,1} = Y_e / \cos \varphi_d\,.
\end{equation}
We can get phenomenologically acceptable couplings only if $\varphi_u \neq
\varphi_d$. 

Note that the high--scale couplings $Y_{f,1}$ are always larger than or equal
to the low--scale (MSSM) couplings $Y_f$ $(f = u,d,e)$. On the other hand, we
know that in the MSSM the top Yukawa coupling is already fairly close to its
upper bound imposed by the requirement that it remains perturbative up to very
large scales. Eq.(\ref{match}) therefore implies that $|\cos \varphi_u|\simeq
1$. For definiteness we therefore set
\begin{equation} \label{phi_u}
\cos \varphi_u = 1\,,
\end{equation}
i.e. $\varphi_u = 0$. This minimizes $Y_{u,1}$; we will see shortly that it
also minimizes all other MSSM matter Yukawa couplings above scale $M_2$. These
couplings appear with positive signs on the right--hand side of the RGE for
the new coupling $Y_N$. The choice (\ref{phi_u}) therefore maximizes the upper
bound on $Y_N(M_R)$ that can be derived from the requirement that this
coupling remains perturbative up to $M_X$. We will see below that this in turn
minimizes the lower bound on the mass of the light physical neutrino for fixed
$M_R$.

Eq.(\ref{Y_3122}) implies that $Y_{u,1}(M_R) = Y_{d,1}(M_R) \equiv
Y_{q,1}(M_R)$. This is compatible with Eqs.(\ref{match}) and (\ref{phi_u})
only for
\begin{equation} \label{phi_d}
\cos \varphi_d = \frac {Y_d(M_2)} {Y_u(M_2)} \left[ \frac {g_1^2(M_R)}
{g_1^2(M_2)} \right]^{1/60} \,;
\end{equation}
the last factor in Eq.(\ref{phi_d}) accounts for the different RGE running of
$Y_{u,1}$ and $Y_{d,1}$ caused by the different hypercharges of the $U^c$ and
$D^c$ superfields. Since this factor is quite close to unity,
Eqs.(\ref{match})--(\ref{phi_d}) imply that $Y_{d,1}$ and $Y_{u,1}$ are very
similar. Since even in the MSSM the bottom and tau Yukawa couplings become
similar at large energy scales, Eq.(\ref{phi_d}) implies that all third
generation Yukawa couplings will be comparable to the top Yukawa coupling at
all scales above $M_2$. In the given framework this is inescapable, unless we
introduce additional heavy superfields which mix with the MSSM matter fields.

At the $SU(4)_C$ breaking scale $Y_{q,1}$ and $Y_{l,1}$ are unified into the
single coupling $Y_1$. The unification of the bottom and top coupling
can always be achieved through an appropriate choice of $\varphi_d$; however,
the unification of the bottom and tau Yukawa couplings is a nontrivial
constraint. This prediction is similar to that of minimal $SU(5)$. In a
scenario without intermediate scales, the tau Yukawa coupling at scale $M_X$
is typically a bit larger than the bottom coupling. In our case unification
should happen at $M_C < M_X$, which reduces the difference between the two
couplings at their putative unification scale. On the other hand, above the
scale $M_1 < M_C$ the $SU(3)_C$ coupling is larger in our scenario than in the
MSSM. This increases the RG running of $Y_b$. The two effects largely
cancel. As a result, we find that $Y_\tau(M_C)$ exceeds $Y_b(M_C)$ by
typically 10 to 20\%. We blame this on threshold effects -- Table 1 shows that
quite a few new fields attain masses of order $M_C$ -- and/or on the
additional physics required to reproduce masses and mixing angles of the
lighter SM fermions. As a practical matter, we set
\begin{equation} \label{Y_q}
Y_1(M_C) = \frac {Y_{l,1}(M_C) + Y_{q,1}(M_C)} {2} \,.
\end{equation}

The superpotential (\ref{Y_3122}) generates neutrino masses through the
celebrated (``type I'') see--saw formula \cite{minkowski},\footnote{Note that
  there is no quartic scalar coupling which could lead to a ``type II'' seesaw
  contribution.}
\begin{eqnarray}\label{eq:seesaw}
m_\nu = \frac{m_D^2}{M_N} = \frac{(Y_{l,1} \langle H_u^0 \rangle)^2}
{Y_N \langle \overline{\sigma} \rangle}\,.
\end{eqnarray}
Here $\overline{\sigma} \in (\overline{\bf 10},{\bf 1,3}) \in
\overline{\Sigma}$ is the neutral component of the $SU(2)_R$ triplet Higgs
boson.\footnote{The field $\bar \delta^0$ listed in Table~1 is the physical
  remnant of $\overline{\sigma}$ after $G_{3122} \rightarrow G_{\rm SM}$
  symmetry breaking.} Note that the neutrino Dirac Yukawa coupling is related
to that of charged leptons by $SU(2)_R$, which in turn is related to the (top)
quark Yukawa coupling by $SU(4)_C$ symmetry, as described above. We assume
here that the $Y_N$ are (mildly) hierarchical, so that only the third
generation coupling is large enough to effect the weak--scale spectrum
significantly via the RGE.

For given $M_R = \langle \overline{\sigma} \rangle$ and light neutrino mass
$m_\nu$, Eq.(\ref{eq:seesaw}) can then be used to determine the value of $Y_N$
at the GUT scale. We vary $m_\nu$ between 0.2 and 0.4 eV. Note that smaller
values of $m_\nu$ lead to a {\em larger} coupling $Y_N$.

The occurrence of fields that are not part of the MSSM at mass scales well
below $M_R$ is crucial. As well known, in the MSSM all three gauge couplings
(almost) meet at an energy scale near $2 \cdot 10^{16}$ GeV
\cite{mssmuni}. Without additional fields that are lighter than $M_R$ it would
not be possible to modify the running of the gauge couplings such that
intermediate scales, and hence energy ranges where the symmetry group is
larger than $G_{\rm SM}$ but smaller than the GUT group, can occur. We will
analyze the running of the gauge couplings in more detail in Sec.~4.

We will see in Sec.~4 that the lightest new particles, with mass $M_2 \sim
M_R^2/M_X$, are still much too heavy to directly lead to visible effects at
collider or rare decay experiments. Nevertheless their existence affects the
renormalization group equations (RGE) describing the running of the masses of
all superparticles and Higgs bosons. The one--loop RGE for Yukawa couplings
and soft breaking parameters that hold for different ranges of energies are
listed in the Appendix. In order to compare with the frequently studied
\cite{cmssm} mSUGRA or cMSSM scenario, we assume universal boundary
conditions, as already noted in the Introduction. 

\subsection{The numerical calculation}

The RGE listed in the Appendix are too complicated to allow an analytical
solution. Instead, we incorporated them into the code {\tt SOFTSUSY2.0}
\cite{allanach}. This program computes the weak--scale MSSM spectrum by
iteratively solving the RGE, starting from universal boundary conditions for
the soft breaking parameters. An iterative treatment is necessary since many
parameters are fixed at the weak scale, rather than the GUT scale. These
include the three (MS)SM gauge couplings, the masses of SM matter
fermions\footnote{Only the masses of third generation fermions are kept, since
  the Yukawa couplings in the first and second generation are too small to
  significantly affect the evolution of the sparticle masses.}, the mass of
the $Z$ boson, and the ratio $\tan\beta$ of vevs of the two MSSM Higgs bosons.
We use one--loop RGE throughout, but include important weak--scale threshold
corrections; these are known to change the physical masses of third generation
fermions significantly, in particular at large $\tan\beta$ \cite{thresh}. Note
that the program implements radiative breaking of the electroweak gauge
symmetry \cite{radbreak}, again including important weak--scale threshold
corrections.

At the intermediate scales $M_1, \, M_2, \, M_R$ and $M_C$ (some of) the RGE
have to be changed. In the discussion of Yukawa couplings we described how to
pick the appropriate high--scale couplings, given the low--energy couplings.
This procedure is applicable when going from low to high energies. When going
in the opposite direction, we use the same matching conditions, employing the
values of $\cos \varphi_d$ and the ratio of $Y_{l,1}(M_C) / Y_{q,1}(M_C)$
determined from the previous RG running from low to high energies to fix the
values of low--scale Yukawa couplings. The matching of gauge couplings and
soft breaking terms directly follows from the group structure, and will be
discussed in Sec.~4.1 and in the Appendix, respectively

The output of {\tt SOFTSUSY} is passed on to the program {\tt micrOMEGAs
  1.3.7} \cite{bbps}, which computes the Dark Matter (DM) relic density as
well as the BR($b\to s\gamma$) and $\delta a_\mu$, the anomalous magnetic
moment of the muon (see below).

\setcounter{footnote}{0}
\section{Accelerator and Cosmological Constraints}

In this Section we describe the constraints we impose on the model.

\subsection{Electroweak symmetry breaking and tachyons}

As mentioned earlier, electroweak symmetry breaking (EWSB) is incorporated
into {\tt SOFTSUSY}. Technically, it solves equations that allow to express
$\mu^2$ and the bilinear Higgs soft mass parameter $B \mu$ in terms of $M_Z$
and the ratio of vevs $\tan\beta$. However, these equations sometimes formally
lead to $\mu^2 < 0$, which indicates that EWSB is not possible for the given
set of input parameters. For reasons that will become clear shortly, here we
are mostly interested in solutions with large $\tan\beta$. In this case EWSB
is possible iff the (properly threshold corrected) value of the squared soft
breaking mass of the up--type Higgs boson at the weak scale is negative,
$m^2_{H_u}(M_{\rm SUSY}) < 0$.

\subsection{Constraints from collider searches}

As in mSUGRA, the most important constraints are those on the masses of the
lightest Higgs boson and the lightest chargino. In combination, they imply
that constraints on the masses of strongly interacting sparticles \cite{pdg}
are automatically satisfied. 

We interpret the limit $M_{H_{\rm SM}} > 114.4$ GeV, which comes from searches
for $e^+ e^- \rightarrow Z H^0$, as imposing a lower mass on the mass of the
lighter CP--even Higgs boson of the MSSM,
\begin{equation} \label{hbound}
M_h > 111 \textrm{ GeV} \, ,
\end{equation}
where we allowed for a $\sim 3$ GeV theoretical uncertainty \cite{herror} in
the calculation of $m_h$. We also require
\begin{equation} \label{cbound}
m_{\tilde \chi_1^\pm} > 104 \ {\rm GeV}\,,
\end{equation}
since scenarios allowing chargino masses significantly below the highest LEP
beam energy cannot be realized in our scenario: these scenarios all require
the presence of sneutrinos with mass near or slightly below that of the
chargino, and scenarios where both the sneutrino and $\tilde \chi_1^\pm$ have
mass below the limit (\ref{cbound}) violate the Higgs constraint
(\ref{hbound}). 

\subsection{Branching ratio of $b \to s\gamma$}

In the SM, flavor changing neutral currents (FCNC) are absent at tree level.
Thus, the radiative $B \to X_s\gamma$ decay is mediated by loops containing
up--type quarks and $W$ bosons. As well known \cite{bsgold}, SUSY loop
contributions can be comparable to those from the SM. Therefore, the
measurement of the branching ratio for this decay, performed by CLEO, Belle
and BaBar \cite{hfag},
\begin{equation} \label{bsg}
B(b \to s\gamma) = (355 \pm 24 ^{+9}_{-10} \pm 3) \times 10^{-6}
\end{equation}
can be used to constrain the parameter space of our model. The first error in
(\ref{bsg}) includes statistical, systematic, extrapolation and $b \to
d\gamma$ contamination errors, while the last two are estimated to be the
difference of the average after varying the central value of each experimental
result by $\pm1\sigma$.  To be conservative, we take the linear sum of the
errors, since the calculation strongly depends on the assumptions of the
boundary conditions. Even minor deviations from strict universality, for
example due to the running between $M_X$ and $M_{Pl}$ \cite{bghw,ekrww}, can
have very large effects \cite{or} .

As mentioned above, we used {\tt micrOMEGAs 1.3} \cite{bbps} to calculate the
branching ratio.  Therein, minimal flavor violation (i.e. the only source
of flavor violation at the weak scale is in the CKM matrix) is assumed
\cite{dgg}; hence only contributions from charged Higgs and top quarks, and
charginos and stops are included. These contributions are indeed usually by
far the dominant ones if universal boundary conditions are assumed
\cite{bbmr}, as in our analysis.

\subsection{The anomalous magnetic moment of the muon}

The anomalous magnetic moment of the muon is one of the most precisely
calculated and measured quantities. There is an about $3 \sigma$ discrepancy
between the SM prediction based on data from $e^+e^-$ annihilation into
hadrons and the experimental value. While this is still somewhat controversial
-- an SM prediction which instead makes use of $\tau$ decay data plus some
assumptions is in fair agreement with the data -- we here want to investigate
the parameter space of our model that allows to explain this discrepancy.

The world average, dominated by data from the E821 collaboration at BNL, is
\cite{pdg}
\begin{equation} \label{amuexp}
a_\mu^{exp} = \frac{g_\mu - 2}{2} = (1165920.80 \pm 0.63) \times 10^{-9} \, .
\end{equation}
The theoretical value \cite{mrr} is calculated as the sum of (i) pure QED
contributions including the diagrams of virtual photon, vacuum polarization
(VP) from $e, \mu$ and $\tau$, and leptonic light--by--light scattering, (ii)
hadronic contributions including VP from quarks, most reliably estimated using
$e^+ e^- \rightarrow$ hadrons data, and hadronic light--by--light scattering,
and (iii) electroweak contributions. The resulting SM prediction is \cite{pdg}
\begin{eqnarray}
a_\mu^{theory} = (1165919.52 \pm 0.52) \times 10^{-9} \, .
\end{eqnarray}
Demanding that supersymmetric loops, involving smuons and neutralinos or smuon
neutrinos and charginos, lead to agreement between theory and experiment at
the $2\sigma$ level thus implies
\begin{equation} \label{amususy}
4.6 \times 10^{-10} < \delta a_{\mu,{\rm SUSY}} < 21.0 \times 10^{-10}\,.  
\end{equation}
We use {\tt micrOMEGAs} to calculate $\delta a_{\mu,{\rm SUSY}}$.

\subsection{Dark Matter relic density}

We assume that all cosmological Dark Matter (DM) consists of lightest
neutralinos. This implies that $\tilde \chi_1^0$ has to be the lightest
superparticle (LSP); this imposes a constraint on the parameter space of our
model.

Far more important is the requirement that the thermal $\tilde \chi_1^0$ relic
density, calculated using {\tt micrOMEGAs} under the usual assumptions of the
minimal cosmological model \cite{kt}, reproduces the value derived from the
WMAP 3-year data \cite{wmap} and other observations pertaining to structure
formation in the universe:
\begin{eqnarray} \label{wmap}
0.097 < \Omega_{\rm DM} h^2 < 0.113,\ \textrm{ at 68\% CL} \,.
\end{eqnarray}
Here $\Omega_{\rm DM}$ is the DM mass density in units of the critical
(closure) density, and $h$ is the Hubble constant in units of 100
km/(s$\cdot$Mpc). As we will see below, this provides the most stringent
constraint on the parameter space. This is not surprising, given the small
size of the error bars in (\ref{wmap}).\footnote{Recently the WMAP 5 year data
  have been released \cite{wmap5}. The resulting range for $\Omega_{\rm DM}
  h^2$ is very similar to that of Eq.(\ref{wmap}).}

\section{Results}

We are now ready to present some results. We begin with an analysis of the
running of the gauge couplings, which determines the values of our
intermediate scales. We then discuss analytical results for first and
second generation sfermion as well as gaugino masses, before analyzing the
ratios of (s)particle masses that are most relevant for the calculation of the
DM relic density. We will conclude this Section with a survey of the parameter
space of the model.

For the top quark mass, we have taken $m_t = 170.9$ GeV, as has recently been
measured at the Tevatron \cite{d0cdf}.

\subsection{RG Analysis of the gauge couplings}

The current world averages of the gauge coupling constants at scale $M_Z$ are
\cite{pdg}:
\begin{eqnarray} \label{coupl}
\alpha_1(M_Z) = 0.01695, \quad \alpha_2(M_Z) = 0.03382, \quad
\alpha_3(M_Z) = 0.1176\,.
\end{eqnarray}
Note that we use GUT normalization for the $U(1)$ gauge coupling, i.e. our
$\alpha_1$ exceeds the hypercharge coupling $\alpha_Y$ in its usual
normalization by a factor of 5/3. The values of these couplings at different
energies are determined by RGE; to one--loop order, these can be written as
\begin{equation} \label{gaugerge}
\frac {d \alpha_i} {d t} = - \frac{\alpha_i} {2 \pi} b_i, \ \ (i=1,2,3)\,.
\end{equation}
Here $t = \ln (Q / Q_0)$, where $Q_0$ is some reference energy scale. Note the
minus sign in Eq.(\ref{gaugerge}); in this convention, a positive $b_i$
corresponds to an asymptotically free gauge coupling. The values of the $b_i$
depend on which particles are ``active'' at a given energy scale $Q$; in the
usual step function approximation of integrating out heavy particles, we treat
all particles with masses $< Q$ to be (fully) active at scale $Q$.

This leads to the values of the $b_i$ listed in Table~\ref{table:beta}, which
we adapted from ref.\cite{aulakh}. Note that we list the coefficients that
allow to describe the running of the three factor groups of the SM gauge
group. The $SU(2)_L$ factor remains independent up to scale $M_X$, i.e. the
third column of Table~\ref{table:beta} always describes the running of the
coupling of an $SU(2)$ group. Recall that above $M_C$, $SU(2)_L$ and $SU(2)_R$
have the same coupling, since the discrete symmetry $D$ is exact; the
coefficient $b_2^{(5)}$ therefore also describes the running of the $SU(2)_R$
coupling. Moreover, at scale $M_C$ the strong interactions get embedded into
$SU(4)_C$, with boundary condition $g_3(M_C) = g_4(M_C)$. The coefficient
$b_3^{(5)}$ therefore describes the running of the $SU(4)_C$ gauge coupling,
which is the same as the running of the coupling of the $SU(3)_C$ subgroup of
$SU(4)_C$ at $Q \geq M_C$.

The fate of the $U(1)_Y$ factor of $G_{\rm SM}$ is a bit more complicated. At
scale $M_R$ it gets embedded into $SU(2)_R \times U(1)_{B-L}$, with matching
condition $\alpha_1^{-1} = 2 / (5 \alpha_{B-L}) + 3 / (5 \alpha_{2R})$. The
$U(1)_{B-L}$ factor in turn gets absorbed into $SU(4)_C$ at scale $M_C$,
i.e. $\alpha_{B-L}(M_C) = \alpha_4(M_C)$. Although the hypercharge coupling is
thus ``spread'' over two different gauge couplings for $Q \geq M_R$, its
running can still be described by Eq.(\ref{gaugerge}), with coefficient listed
in Table~\ref{table:beta}.

\begin{table}[!tp]
\caption{The coefficients of the beta functions of the gauge couplings of the
  SM gauge group, valid at different energy scales $Q$.}
\begin{center}
\begin{tabular}{c|c|c|c}
Energy range & $b_1^{(k)}$ & $b_2^{(k)}$ & $b_3^{(k)}$ \\
\hline
$M_Z < Q < M_S$ & $-41/10$  & $ 19/6$ & $7$ \\
$M_S < Q < M_2$ &$ -33/5 $ &  $ -1$ & $ 3$  \\
$M_2< Q < M_R$ &$ -12$ & $-2$ &
$3$\\
$M_R< Q <M_1$  & $-48/5$ & $-2$ &
$3$ \\
$M_1 < Q < M_C $&$ -48/5$ & $-2$ &$0$ \\
$M_C < Q < M_X $& $-194/5$ & $-42$ & $-34$\\
\end{tabular}
\end{center}
\label{table:beta}
\end{table}

Eqs.(\ref{coupl}) and (\ref{gaugerge}), together with the coefficients
$b_i^{(k)}$ listed in Table~\ref{table:beta}, allow us to predict the values
of the gauge couplings at all $Q \geq M_Z$. Of course, the three gauge
couplings of the (MS)SM are supposed to meet at scale $M_X$ in our model. This
leads to two independent constraints. On the other hand, the intermediate
scale $M_R$ and $M_C$ are free parameters of our model; the scales $M_1$ and
$M_2$ are derived quantities, as described in Table~\ref{table:mass}. For
given value of $M_X$ the two independent unification conditions can thus be
solved for $M_R$ and $M_C$. The running of any one of the three (MS)SM gauge
couplings can then be used to determine the value of the $SO(10)$ gauge
coupling $\alpha_U$. Notice that this procedure will work for any assumed
value of $M_X$, i.e. it still leaves one parameter undetermined. We refer the
reader to ref.\cite{aulakh} for a further discussion of the unification
condition, including explicit solutions of the RGE of the gauge couplings.

\begin{figure}[!bp]
\begin{center}
\includegraphics[width=8.2cm,height=8.2cm,angle=270]{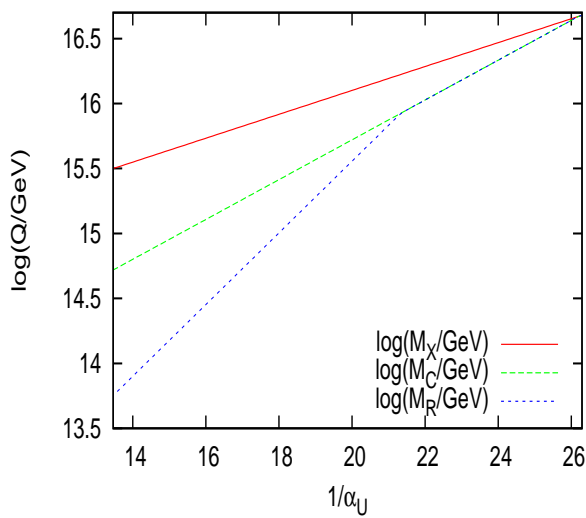}
\caption{The values of the intermediate scales $M_R$ and $M_C$, as function of
the inverse of the $SO(10)$ coupling $\alpha_U$. Here we took $M_S = 1$ TeV as
sparticle mass scale.}  
    \label{fig:escale}
\end{center}
\end{figure}

In Fig.~\ref{fig:escale} we show one--loop predictions for the intermediate
scales $M_R$ and $M_C$, as well as the value of $M_X$, as function of
$1/\alpha_U$. We see that smaller values of $M_X$ correspond to larger values
of $\alpha_U$. The reason is that decreasing $M_X$ increases the ratios
$M_X/M_R$ and $M_X / M_C$. Table~\ref{table:beta} shows that all $b_i$ are
large and negative for $Q > M_C$; recall that this corresponds to gauge
couplings increasing with energy. A large $M_X / M_R$ means that these
beta--functions are valid over a large range of energies, leading to a large
value of $\alpha_U$. On the other hand, proton decay through dimension 6
operators conservatively requires $M_X \geq 3 \cdot 10^{15}$ GeV.
Fig.~\ref{fig:escale} shows that this corresponds to $\alpha_U \simeq 1/13.5$,
safely in the perturbative region (significantly smaller than $\alpha_3(M_Z)$,
for example). 

Since the purpose of our paper is to study the influence of the intermediate
scales on the low--energy spectrum, we take this minimal value of $M_X$ as our
default choice. The intermediate scales are then found at
\begin{equation} \label{mrmc}
M_R = 10^{13.75} \ {\rm GeV}\,, \ \ \ M_C = 10^{14.72} \ {\rm GeV}\,.
\end{equation}
Increasing $M_X$ reduces the impact of the intermediate scales. At $M_X \simeq
10^{15.8}$ GeV, corresponding to $\alpha_U \simeq 1/21$, the scales $M_R$ and
$M_C$ coincide. When $M_X$ is increased to about $10^{16.6}$ GeV, $M_C$ in
turn coincides with $M_X$. At that point no intermediate scales are left,
i.e. this limit reproduces the usual MSSM. Higher values of $M_X$ are not
possible. By varying $M_X$ between $10^{15.5}$ GeV and $10^{16.6}$ GeV we can
thus smoothly turn on the intermediate scales and study their impact on
weak--scale physics.

\begin{figure}[h!]
\begin{center}
\includegraphics[width=8.2cm,height=8.2cm,angle=270]{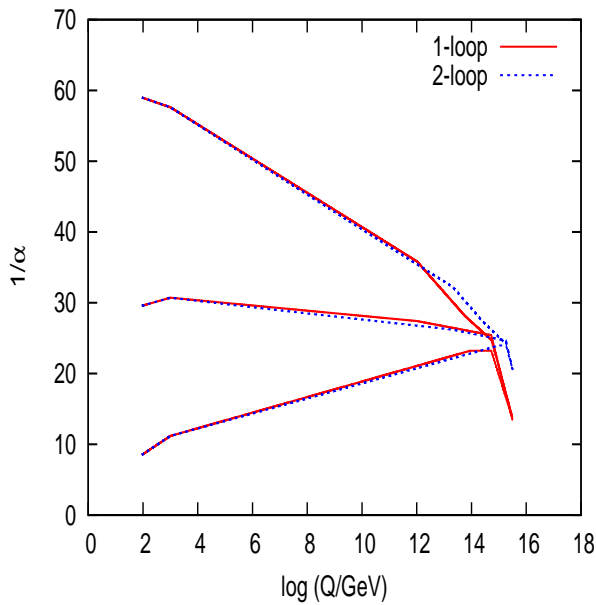}
\caption{The running inverse gauge couplings $\alpha_i^{-1}$. The curves at
    the top [middle, bottom] are for the $U(1)_Y$ [$SU(2)_L, \ SU(3)_C$]
    couplings. Solid and dashed curves show results for one-- and two--loop
    RGE, respectively.}
    \label{fig:gauge}
\end{center}
\end{figure}

The rapid increase of the gauge couplings at $Q \geq M_C$ can also be seen in
Fig.\ref{fig:gauge}, which shows the running of the gauge couplings as
function of the energy scale for our default set of parameters. The solid
lines show the predictions from the one--loop RGE we have used so far, whereas
the dashed curves are based on two--loop RGE \cite{lp} (ignoring, however, the
subdominant contributions from Yukawa couplings to the running of the gauge
couplings). Evidently using two--loop RGE increases the intermediate scales
for this value of $M_X$, making the model more mSUGRA--like. However, an
analysis based on two--loop RGE should also treat the (rather numerous, in our
case) threshold corrections more carefully. So far we have assumed that all
(s)particles whose masses are of the order of a given scale, as listed in
Table~\ref{table:mass}, have {\em exactly} that mass. This will not be true in
many cases. However, the exact masses will depend on many unknown couplings
describing interactions of these superheavy fields. A proper treatment of
threshold corrections would therefore introduce many new free parameters.
Since threshold and two--loop effects are generically of similar magnitude
\cite{lp}, we assume that there are combinations of parameters where an
analysis including two--loop and threshold effects leads to similar results as
the one--loop analysis. The use of one--loop beta functions has the practical
advantage that the equations determining $M_R$ and $M_C$ can easily be solved
analytically \cite{aulakh}.

\setcounter{footnote}{0}
\subsection{Analytical results}

At the one--loop level, the gaugino masses evolve in the same way as the
squared gauge couplings do. Therefore, the ratios of weak--scale gaugino
masses are the same as in mSUGRA, i.e. $M_1:M_2:M_3 \simeq
1:2:6$.\footnote{These are running masses. The on--shell masses differ by
  weak--scale threshold corrections \cite{thresh}, which are included in {\tt
    SOFTSUSY}.} This follows from the fact that the three MSSM gauge couplings
are identical at $M_X$, and have their measured values (\ref{coupl}) at scale
$M_Z$. These ratios are therefore independent of the intermediate scales.

However, for fixed $M_{1/2}$ the weak--scale gaugino masses are now much
smaller than in mSUGRA, since the ratios $\alpha_i(M_Z) / \alpha_U$ are much
smaller, as shown in Fig.~\ref{fig:gauge}. Writing
\begin{equation} \label{gaugino}
M_i(M_{\rm SUSY}) = c_i M_{1/2} \ \ \ \ (i=1\,,2\,,3)\,,
\end{equation}
we have 
\begin{equation} \label{ci}
c_1 \simeq 0.23\,,\ \ c_2 \simeq 0.46\,,\ \ c_3 \simeq 1.4\,,
\end{equation}
for $M_{\rm SUSY} \sim 1$ TeV; these are nearly two times smaller than the
corresponding coefficients in mSUGRA \cite{msugra}.

The RGE for the masses of first and second generation sfermions, whose Yukawa
couplings are negligible, can also be solved analytically \cite{msugra}.
Writing\footnote{The running weak--scale sfermion masses also receive small
  $D-$term contributions, which we omit in the following discussion, but
  include in the numerical analysis. In addition, the physical (pole) masses
  again differ from the running masses by threshold corrections \cite{thresh},
  which are included in {\tt SOFTSUSY}.}
\begin{equation} \label{sfermion}
m_{\tilde f}^2 (M_{\rm SUSY}) = m_0^2 + c_{\tilde f} M_{1/2}^2\,,
\end{equation}
we have
\begin{equation} \label{selsquark}
c_{\tilde e_R} \simeq 0.15\,, \ \ c_{\tilde l_L} \simeq 0.21\,, \ \ 
c_{\tilde q} \simeq 1.16\,.
\end{equation}
Here $\tilde e_R$ and $\tilde l_L$ stands for $U(1)_Y$ singlet and doublet
sleptons, respectively, while $\tilde q$ stands for an average first or second
generation squark; as in mSUGRA, $SU(2)_L$ doublet squarks are slightly
heavier than singlet squarks. We checked that the analytical and numerical
calculations of $m^2_{\tilde e_R}$ match within 0.1$\%$. Note that the
coefficient $c_{\tilde e_R}$ is numerically almost the same as in mSUGRA
\cite{msugra}.  This is due to a cancellation of two effects. On the one hand,
$\tilde e_R$ is a non--singlet under both $SU(2)_R$ and $SU(4)_C$, giving rise
to new gauge contributions to its mass at scales above $M_R$ and $M_C$,
respectively, which increase $c_{\tilde e_R}$. On the other hand, we saw that
for fixed $M_{1/2}$ the gaugino masses at scales $Q < M_X$ are smaller than in
mSUGRA, which reduces all $c_{\tilde f}$. The latter effect is dominant for
all fields that transform non--trivially under either $SU(2)_L$ or $SU(3)_C$.
As a result, the mass difference between $SU(2)_L$ singlet and doublet
sleptons is significantly smaller than in mSUGRA; recall that the $SU(2)_L$
doublet sleptons are singlets under $SU(2)_R$.

Note that if we apply the universal boundary conditions at some energy scale
$Q > M_X$, the sfermion masses obtain additional contributions due to $SO(10)$
gauge interactions. These increase the values of all $c_{\tilde f}$ by the
same amount, since all sfermions reside in the {\bf 16} of $SO(10)$; this
additional contribution would thus be relatively most important for $\tilde
e_R$ \cite{cmv}. However, since we need large Higgs representations to realize
the breaking chain (\ref{chain}), the $SO(10)$ gauge coupling $\alpha_U$ hits
a Landau pole soon after the unification scale \cite{aulakh}. Hence we expect
some new, possibly strongly interacting, physics to occur just above $M_X$.
The range of energies where $SO(10)$ RGE are applicable is therefore probably
quite small.

From Eqs.(\ref{gaugino})--(\ref{selsquark}) we can derive lower bounds on the
ratios of sfermion to gaugino masses. Of particular interest for the
calculation of the Dark Matter relic density is the relation
\begin{equation} \label{selrat}
\frac {m_{{\tilde e}_R}(M_{\rm SUSY})} {|M_1|(M_{\rm SUSY})} \gsim 1.68\,.
\end{equation}
In mSUGRA, the lower bound, which is saturated for $M_{1/2}^2 \gg m_0^2$, is
instead slightly below unity. This is important, since it implies that for
fixed $m_0$ and increasing $M_{1/2}$, one will eventually reach a gaugino mass
such that $m_{\tilde e_R} = m_{\tilde \chi_1^0}$, leading to strong $\tilde
\chi_1^0 - \tilde e_R$ co--annihilation. The bound (\ref{selrat}) implies that
this never happens in our scenario. However, as in mSUGRA the lighter $\tilde
\tau$ mass eigenstate $\tilde \tau_1$ can be significantly lighter than
$\tilde e_R$.\footnote{For this reason, usually the most important
  co--annihilation channel is $\tilde \chi_1^0 - \tilde \tau_1$
  co--annihilation \cite{coann}: scenarios giving $m_{\tilde \chi_1^0} =
  m_{\tilde e_R}$ are already excluded, since here $\tilde \tau_1$ would be
  the LSP.} We will see later that $\tilde \chi_1^0 - \tilde \tau_1$
co--annihilation remains possible in our model. However, Eq.(\ref{selrat})
already indicates that the parameter space where this can happen is (even)
more limited than in mSUGRA.

Our model also predicts
\begin{equation} \label{selrat2}
\frac { m_{\tilde l_L}(M_{\rm SUSY}) } { |M_2(M_{\rm SUSY})| } \gsim 1\,,
\end{equation}
which means that $\tilde \chi_2^0$ and $\tilde \chi_1^\pm$ decays into
$SU(2)_L$ doublet sleptons will be strongly suppressed. In contrast, in mSUGRA
$SU(2)_L$ doublet sleptons can be some 15\% lighter than $SU(2)_L$
gauginos. On the other hand, the bound
\begin{equation} \label{squarkrat}
\frac { m_{\tilde q}(M_{\rm SUSY}) } { |M_3(M_{\rm SUSY})| } \gsim 0.77\
\end{equation}
is very similar to that in mSUGRA. It still leaves room for two--body decays
of gluinos into first or second generation squarks.

\setcounter{footnote}{0}
\subsection{Mass ratios}

In this section we show numerical results for some (ratios of) masses that are
important for the determination of the thermal $\tilde \chi_1^0$ relic
density. We focus on masses whose weak--scale values are affected by the
potentially large Yukawa couplings in the theory. We saw in Sec.~2.1 that all
third generation Yukawa couplings involving Higgs doublets are quite large at
energies $\geq M_2$; at energies $\geq M_R$ this includes the new neutrino
Yukawa coupling $Y_\nu$, which is equal to that of the charged lepton by
$SU(2)_R$ invariance. The coupling $Y_N$, which determines the Majorana masses
of the heavy neutrinos, can also be sizable. Recall that $Y_N$ is related to
the light neutrino mass and $M_R = \langle \bar{\sigma} \rangle$ through
Eq.(\ref{eq:seesaw}).  

Yukawa couplings tend to reduce weak--scale scalar masses for fixed $m_0$ and
$M_{1/2}$. $Y_N$ begins to act -- on the mass of $\tilde \tau_R$ -- at scale
$M_2$; at the same scale, the bottom and tau couplings become large even if
$\tan\beta$ is not large, see Eq.(\ref{phi_d}). At energies above $M_R$ the
neutrino coupling $Y_\nu$ becomes active, reducing the weak--scale masses of
$\tilde \tau_L$ and of the Higgs boson $H_u$. Above $M_C$, all weak--scale
third generation sfermion masses will be reduced by $Y_N$. We therefore expect
the difference between first and third generation weak--scale sfermion masses
to be larger than in mSUGRA. This effect should be strongest for $\tilde
\tau_R$ and $\tilde \tau_L$. The reduction should be more pronounced at small
and moderate $\tan\beta$, since for large $\tan\beta$ all third generation
Yukawa couplings are sizable even in the MSSM.

\begin{figure}[!tp]
\includegraphics[width=8.2cm, height=8.2cm, angle=270]{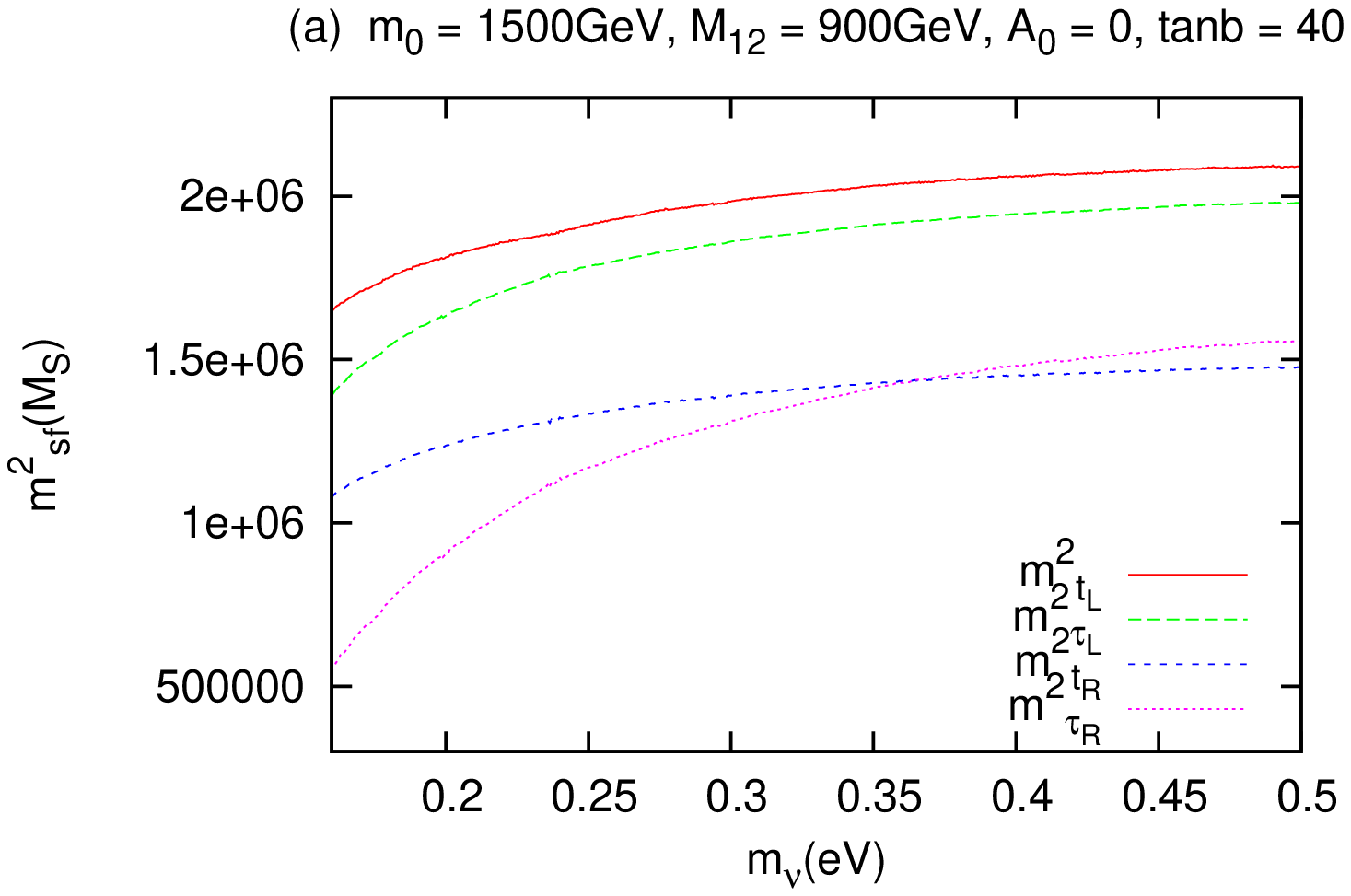}
\includegraphics[width=8.2cm, height=8.2cm, angle=270]{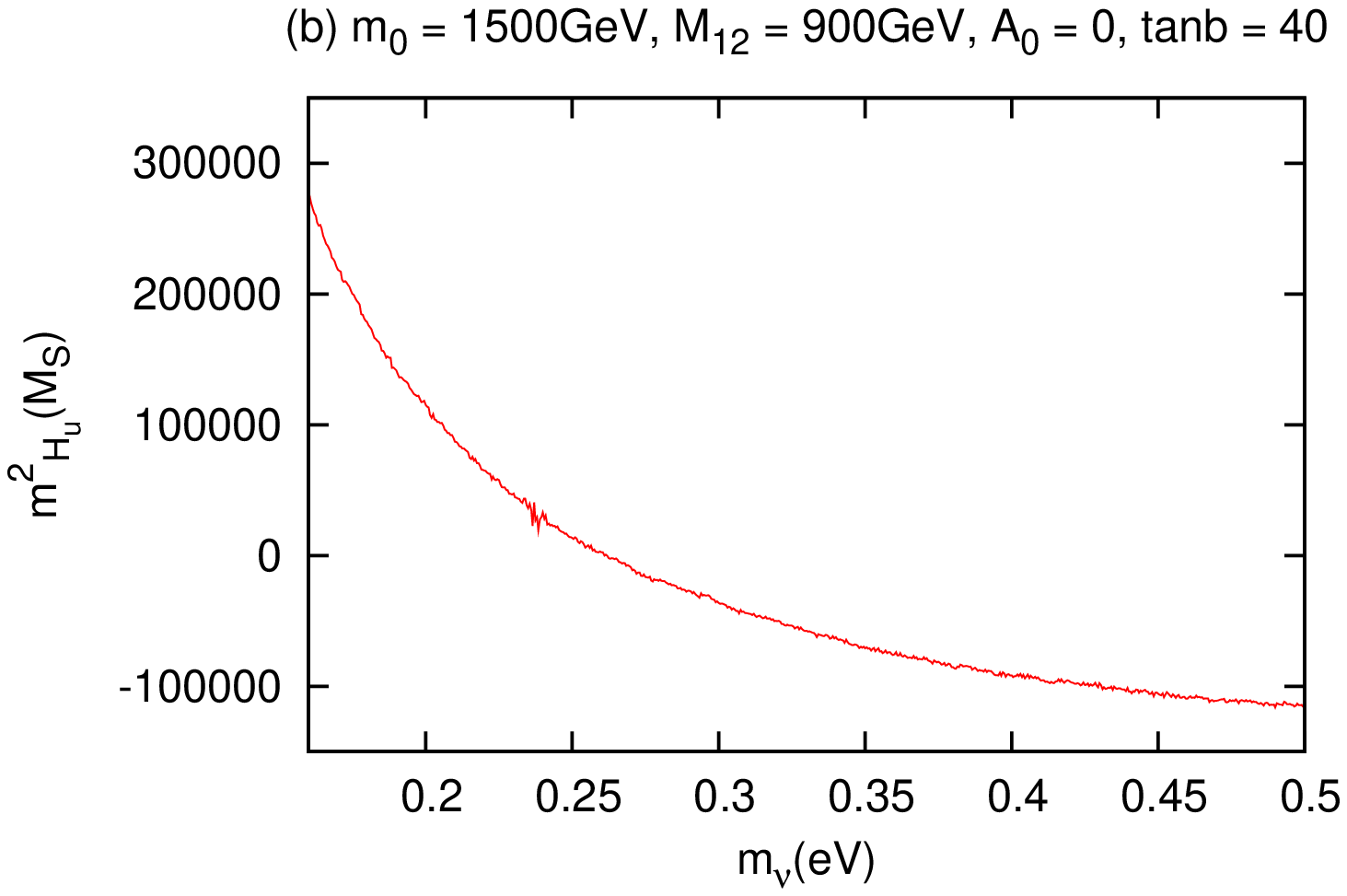}
\caption{Squared weak--scale running masses in GeV$^2$ of (a) $\tilde t$ and
  $\tilde \tau$ sfermions and (b) the up--type Higgs boson, as function of the 
  neutrino mass. The other input parameters are: $m_0 = 1.5$ TeV, $M_{1/2} =
  0.9$ TeV, $A_0 = 0$, $\tan\beta = 40$ and $\mu > 0$.}
    \label{fig:mnu}
\end{figure}

This is illustrated in the left frame of Fig.~\ref{fig:mnu}, which shows the
dependence of the soft--breaking masses of $\tilde t_L, \,\tilde t_R, \,
\tilde \tau_L$ and $\tilde \tau_R$ as a function of the mass $m_\nu$ of the
heaviest light neutrino. Recall that this mass is proportional to $1/Y_N$,
i.e. smaller $m_\nu$ correspond to larger $Y_N$, and hence to smaller
weak--scale sfermion masses.

On the other hand, the right frame in Fig.~\ref{fig:mnu} shows that the
running soft breaking mass of the Higgs bosons with positive hypercharge {\em
  in}creases with decreasing $m_\nu$. We just saw that larger values of $Y_N$
reduce $m^2_{\tilde t_L}$ and $m^2_{\tilde t_R}$ at all energies below $M_X$.
This reduces the term $\propto Y_t^2$. Since this term drives $m^2_{H_u}$ to
smaller or even negative values, reducing its size leads to an increase of the
weak--scale value of $m^2_{H_u}$. Recall that $m^2_{H_u}(M_{\rm SUSY}) < 0$ is
required to achieve electroweak symmetry breaking with $\tan\beta \gg 1$. This
figure therefore implies that the parameter space permitting radiative
breaking of the $SU(2) \times U(1)_Y$ symmetry will be smaller for smaller
values of $m_\nu$.

The Yukawa coupling $Y_N$ also reduces the value of the soft breaking mass of
$\bar \Sigma$, whose vev is responsible for the masses of the heavy neutrinos,
and contributes to the $SU(2)_R \times U(1)_{B-L} \rightarrow U(1)_Y$
breaking. Since this occurs at a scale $M_R \gg M_{\rm SUSY}$, we introduced a
field $\Sigma$, which permits to keep the $SU(2)_R$ and $U(1)_{B-L} \ D-$terms
much below $M^2_R$. However, since $Y_N \neq 0$ implies $m^2_{\bar
  \Sigma}(M_R) < m^2_{\Sigma}(M_R)$, $\langle \Sigma \rangle < \langle
\bar{\Sigma} \rangle$. Since $\langle \bar{\Sigma} \rangle^2 - \langle \Sigma
\rangle^2 \propto (m^2_{\Sigma} - m^2_{\bar{\Sigma}}) \propto M^2_{\rm SUSY}$
\cite{drees}, this effect does not spoil the hierarchy $M_R \gg M_{\rm SUSY}$,
but it does give new non--vanishing contributions to the masses of sfermions
and Higgs bosons.  However, they are subdominant for most of the parameter
space, partly due to the small splitting between $M_R$ and $M_C$, and partly
because $m^2_{\Sigma}$ and $m^2_{\bar \Sigma}$ receive identical, large gauge
contributions, in particular for $Q > M_C$ where $\Sigma$ and $\bar \Sigma$
are embedded in (anti--)decuplets of $SU(4)_C$. In fact, we found that the
ratio $(m^2_{\Sigma} - m^2_{\bar{\Sigma}})/m^2_{\Sigma}$ is, at most, a few
$\%$.  Hence, these new $D-$term contributions can be ignored.

\begin{figure}[!tp]
\includegraphics[width=8.2cm,height=8.2cm, angle=270]{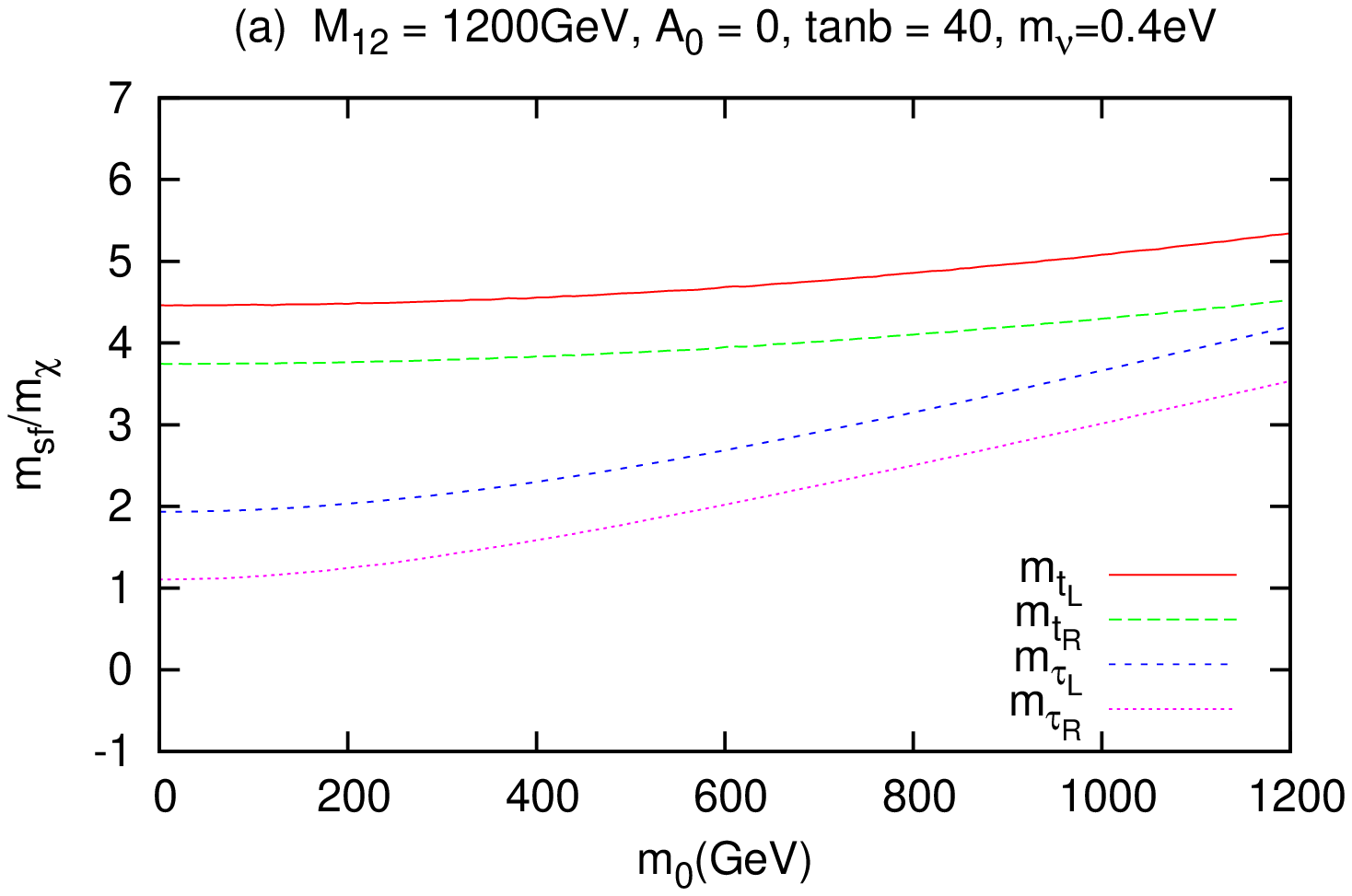}
\includegraphics[width=8.2cm,height=8.2cm, angle=270]{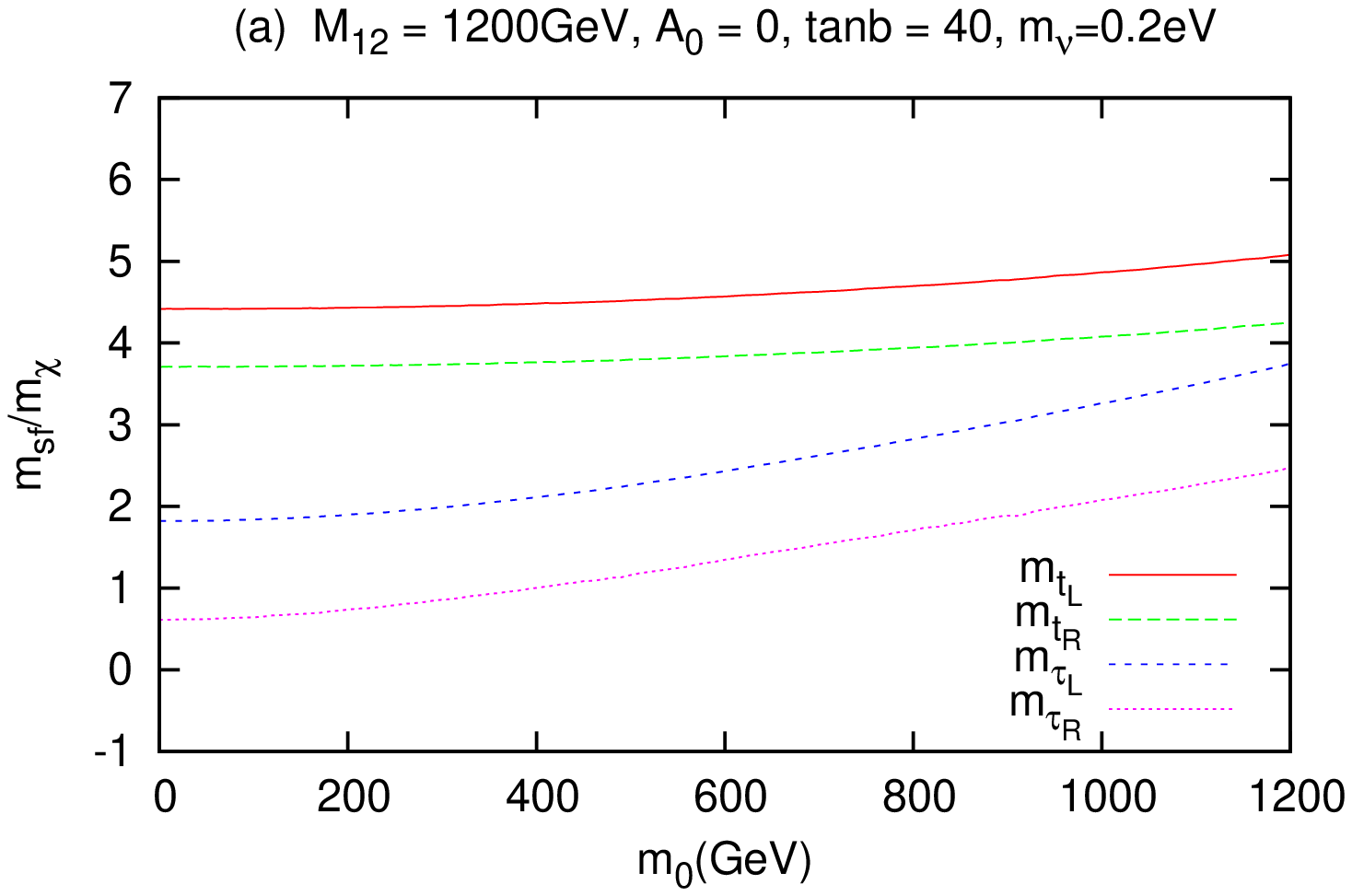}
\caption{The ratio of third generation sfermion masses to the mass of the
  lightest neutralino as function of $m_0$, for $m_\nu =0.4$ eV (left), $m_\nu
  = 0.2$ eV (right). The values of the other input parameters are $M_{1/2} =
  1.2$ TeV, $A_0 = 0,\, \tan\beta = 40$ and $\mu > 0$.} 
    \label{fig:stopstau}
\end{figure}

Fig.~\ref{fig:stopstau} shows the dependence of the ratios $m_{{\tilde
    t}_{L,R}} / m_{\tilde \chi^0_1}$ and $m_{{\tilde \tau}_{L,R}} / m_{\tilde
  \chi^0_1}$, taken at scale $Q = M_{\rm SUSY}$, on $m_0$. For $m_0^2 \ll
M^2_{1/2}$ the stop squarks are significantly heavier than the stau sleptons.
This qualitative behavior is the same as for first and second generation
squarks, see Eq.(\ref{selsquark}). On the other hand, if $m_0 \gsim M_{1/2}$
and relatively small $Y_N$ (left frame), $m_{\tilde t_R}$ can be smaller than
$m_{\tilde \tau_{L,R}}$, since the top Yukawa coupling is significantly larger
than that of the $\tau$ lepton.

The right frame of Fig.~\ref{fig:stopstau} shows that increasing $Y_N$ reduces
the dependence of third generation sfermion masses on $m_0$. In fact, for
$m_\nu = 0.2$ eV we observe a sort of ``focus point'' \cite{focus} for
$m_{\tilde t_R}$, i.e. $m_{\tilde t_R} (M_{\rm SUSY})$ becomes almost
independent of $m_0$. This implies that there is {\em no} focus point behavior
of $m^2_{H_u} (M_{\rm SUSY})$, i.e. this soft breaking parameter, which
largely determines electroweak symmetry breaking for $\tan^2 \beta \gg 1$,
does depend on $m_0$. Hence large values of $m_0$ will not be ``natural'' (by
the definition employed in refs.\cite{focus}) if $Y_N$ affects the weak--scale
third generation masses significantly. 

The scalar masses shown in Fig.~\ref{fig:stopstau} are running masses at scale
$M_{\rm SUSY}$. The physical masses will be affected by threshold corrections
and, more importantly for third generation sfermions, by mixing between
$SU(2)$ singlets and doublets. This mixing reduces the mass of the lighter
eigenstates $\tilde \tau_1$ and $\tilde t_1$, so that $m_{\tilde \tau_1} <
{\rm min}(m_{\tilde \tau_L},\, m_{\tilde \tau_R})$ and similar for $m_{\tilde
  t_1}$. Nevertheless Fig.~\ref{fig:stopstau} shows that co--annihilation will
usually only be possible with $\tilde \tau_1$. This is similar to
mSUGRA. Note, however, that Fig.~\ref{fig:stopstau} is for $\tan\beta =
40$. Recall that co--annihilation with first or second generation sfermions is
not possible here, see Eq.(\ref{selrat}). Moreover, comparison of the two
frames of Fig.~\ref{fig:stopstau} shows that the effects of $Y_N$ on the
$\tilde \tau$ masses are quite small if $m_0^2 \ll M_{1/2}^2$. As a result, we
find that $\tilde \tau$ co--annihilation is possible in our model only for
$\tan\beta \gsim 27$.

\begin{figure}[!tp]
\includegraphics[width=8.2cm,height=8.2cm,angle=270]{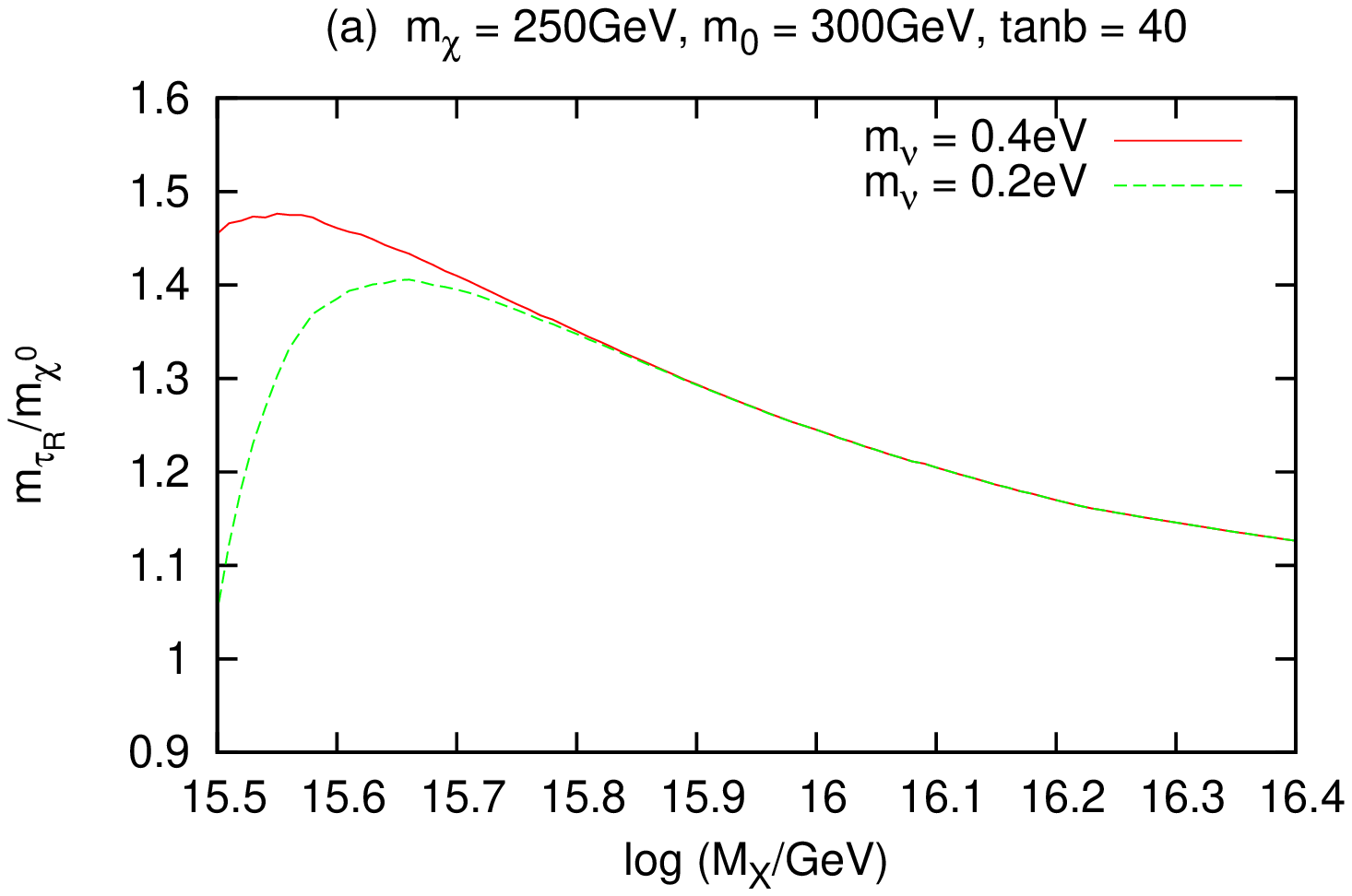}
\includegraphics[width=8.2cm,height=8.2cm,angle=270]{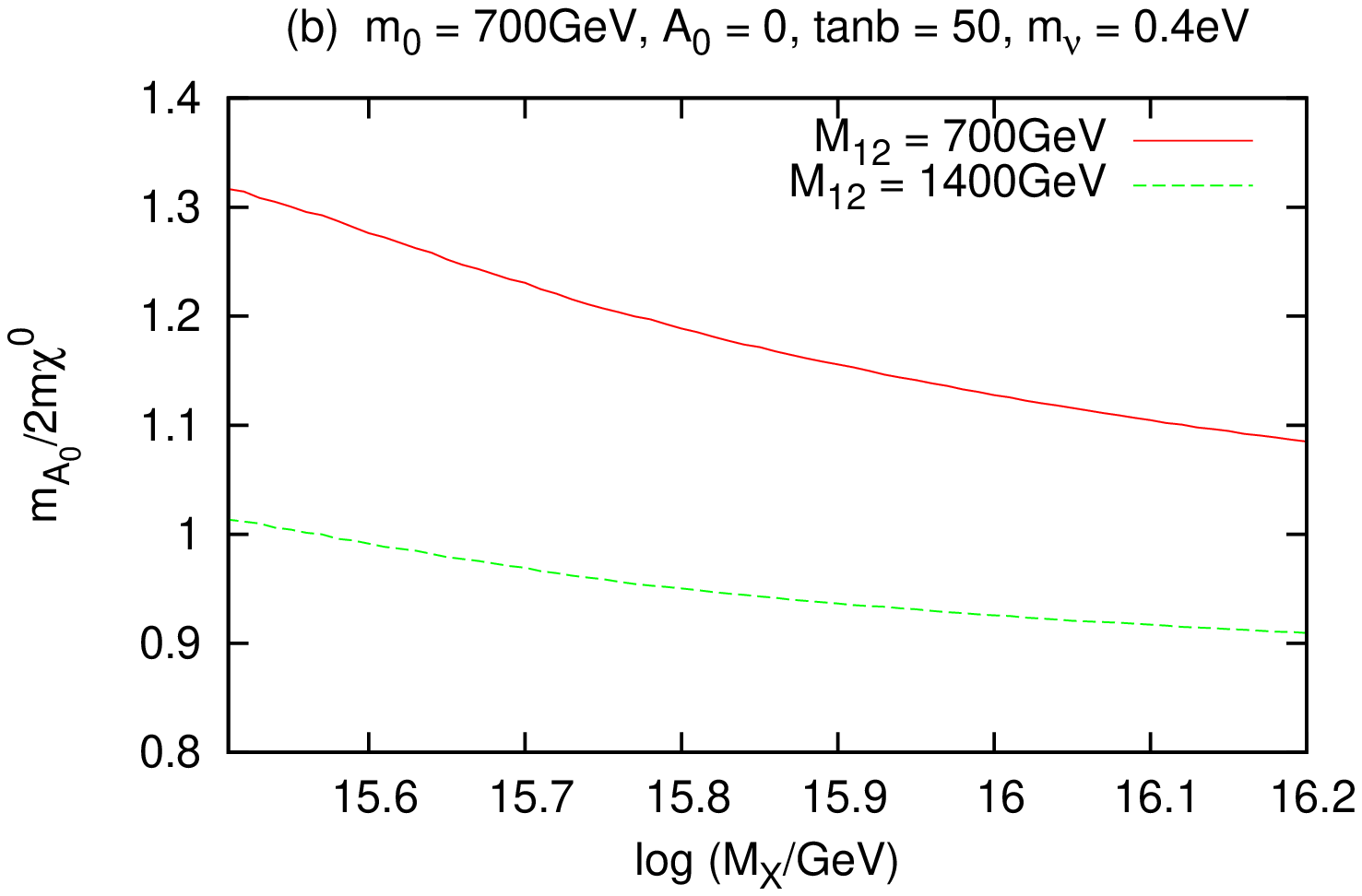}
\caption{The ratio of $m_{\tilde{\tau_R}}$ to $m_{\tilde{\chi}^0}$ (left),
    $m_{\tilde{A^0}}$ to $2m_{\tilde{\chi}^0}$ (right), as the unification
    scale $M_X$ is varied. Recall that this implies corresponding variations
    of the intermediate scales $M_C$ and $M_R$, see Fig.~\ref{fig:escale}.} 
    \label{fig:mx}
\end{figure}

The left frame of Fig.\ref{fig:mx} illustrates the dependence of the ratio
$m_{{\tilde{\tau}}_R}/m_{\tilde{\chi}^0}$ on the GUT scale. Recall that for
$M_X = 10^{16.4}$ GeV our model becomes indistinguishable from mSUGRA, as
far as the weak--scale spectrum is concerned. As discussed in the previous
Subsection, in the absence of new large Yukawa couplings this ratio can only
become larger as the intermediate scale is turned on.  However, we saw in
Fig.\ref{fig:stopstau} that the Majorana Yukawa coupling $Y_N$ does give a
large positive contribution to the RGE of $m_{\tilde \tau_R}$, reducing its
weak--scale value.  These two effects clearly compete with each other.  We see
that even a rather large $Y_N$, corresponding to $m_{\nu_\tau} = 0.2$ eV, can
change $m_{\tilde \tau_R}(M_{\rm SUSY})$ significantly only if $M_X <
10^{15.8}$ GeV; recall from Fig.~\ref{fig:escale} that this corresponds to the
region of parameter space where $M_R < M_C$. In this case the possibility to
have $\tilde \tau_1$ co--annihilation obviously strongly depends on
$m_{\nu_\tau}$.

The right frame of Fig.\ref{fig:mx} shows the ratio $m_{A^0} /
2m_{\tilde{\chi}^0}$. This ratio needs to be close to unity for $\tilde
\chi_1^0$ annihilation through $s-$channel $A^0$ exchange to be enhanced. We
see that reducing $M_X$, i.e. turning on the intermediate scales, slightly
increases this ratio even if $Y_N$ is small. For given $m_0$, this can be
compensated by increasing $M_{1/2}$. We thus expect the ``$A-$funnel'' region
to survive in our scenario, if $\tan\beta \gsim 50$ and for small $Y_N$.
Increasing $Y_N$ will increase $m_A(M_{\rm SUSY})$; this is analogous to the
increase of $m^2_{H_u}$ depicted in the right frame of Fig.~\ref{fig:mnu}. 

\subsection{Regions of the $(m_0,\, M_{1/2})$ plane}

In this Subsection, we show the $(m_0, M_{1/2})$ plane of our model,
indicating the regions where the various accelerator as well cosmological
constraints discussed in Sec. 3 are satisfied.  We scan the parameter space
only up to $(m_0, M_{1/2}) =$ (2000 GeV, 1500 GeV). Even larger sparticle
masses appear quite unnatural. The LHC should be able to probe the entire
parameter space we show \cite{lhc}; recall that $M_{1/2} = 1.5$ TeV
corresponds to a gluino mass around 2 TeV in our scenario. We focus on large
values of $\tan\beta$. We saw in the previous subsection that this is required
both for the $A-$funnel and the for $\tilde \tau$ co--annihilation region in
our scenario. Finally, sign($\mu$) is chosen positive in all plots, in
accordance with the indication of an additional positive contribution to
$g_\mu$; recall also that taking $\mu > 0$ makes it easier to satisfy the $b
\rightarrow s \gamma$ constraint \cite{bbmr}.

\begin{figure}[!tp]
\includegraphics[width=8.2cm,height=8.2cm,angle=270]{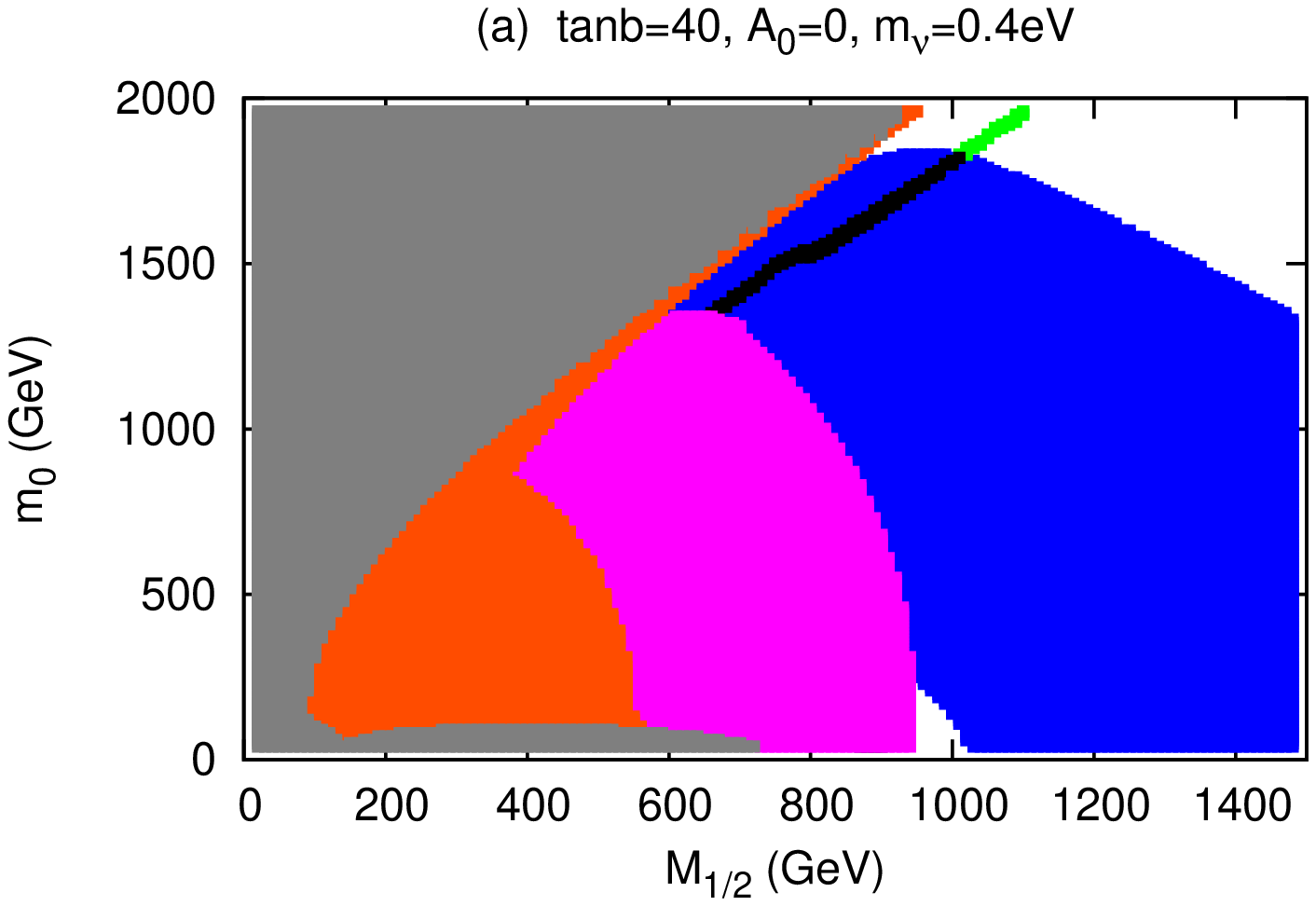}
\includegraphics[width=8.2cm,height=8.2cm,angle=270]{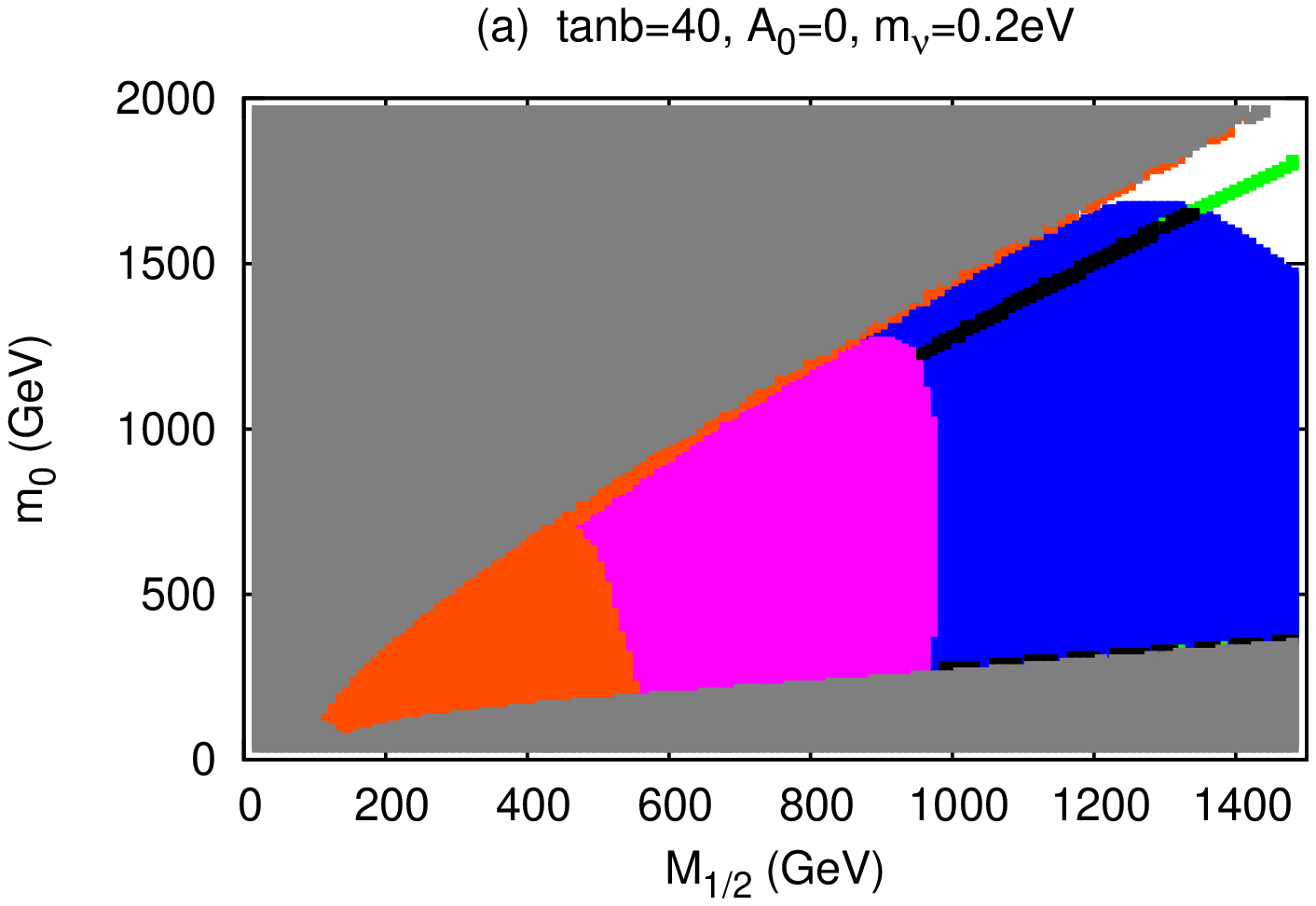}
\caption{Constraints on the $(m_0, M_{1/2})$ plane of our model. The grey
    areas are those excluded by the EWSB condition or by tachyonic or too light
    sfermions. The region excluded by the Higgs and chargino mass constraints
    is shown in bright red and the own excluded by the  $b \to s\gamma$
    constraint in pink. The blue area satisfies the $g_\mu -2$ constraint
    (\ref{amususy}), while green regions satisfy the Dark Matter constraint
    (\ref{wmap}). Finally, black regions satisfy all constraints.}
    \label{fig:tanb40}
\end{figure}

A first example, for $A_0 = 0$ and $\tan\beta = 40$, is presented in
Fig.~\ref{fig:tanb40}; the left (right) frame is for small (large) coupling
$Y_N$. The grey regions are mostly excluded by the requirement of correct
electroweak symmetry breaking; in the right frame the region of small $m_0$ is
instead excluded because $\tilde \tau_1$ is too light (below either the LEP
limit or the mass of $\tilde \chi_1^0$). As expected from the discussion of
Fig.~\ref{fig:mnu}, this region is considerably larger for large $Y_N$. 

The bright red regions are excluded by the chargino search limit
(\ref{cbound}) or by the limit (\ref{hbound}) on the mass of the lightest
CP--even Higgs boson; the latter is relevant for $M_{1/2} \lsim 500$ GeV,
while the former excludes the narrow red strip bordering the grey region at
large $m_0$ and large $M_{1/2}$. Finally, the pink regions are excluded by the
constraint (\ref{bsg}) on the branching ratio for radiative $b$ decays. Some
supersymmetric contributions to the corresponding amplitude grow $\propto \tan
\beta$. This constraint therefore becomes relevant at the large values of
$\tan\beta$ required to realize $\tilde \tau$ co--annihilation and/or the
$A-$funnel in our model. 

Turning to observables that require a non--vanishing contribution from
supersymmetric particles, in the blue regions the constraint (\ref{amususy})
from the anomalous magnetic moment of the muon is satisfied. The corresponding
diagrams are quite similar to those contributing to $b \rightarrow s \gamma$
decays. In particular, some contributions again grow $\propto \tan\beta$. As
in mSUGRA \cite{cmssm}, we find regions of the parameter space at sufficiently
large $M_{1/2}$ where electroweak gauginos and sleptons are sufficiently light
to give a sizable positive contribution to $g_\mu$, while (stop) squarks are
sufficiently heavy not to reduce the branching ratio for $b \rightarrow s
\gamma$ decays too much.

Note that the red, pink and blue regions all extend to much larger values of
$M_{1/2}$ than in mSUGRA \cite{cmssm}. The reason is that the corresponding
constraints probe weak--scale (s)particle masses; we saw in
Eqs.(\ref{gaugino})--(\ref{selsquark}) that a given $M_{1/2}$ corresponds to
much lighter gauginos and sfermions in our scenario than in mSUGRA. Moreover,
we saw in Figs.~\ref{fig:mnu} and \ref{fig:stopstau} that the additional large
Yukawa couplings in our model tend to reduce weak--scale stop masses. They
also increase $m^2_{H_u}$, which leads to a reduction of $|\mu|$ via the
condition of electroweak symmetry breaking. Both effects, which become more
important for smaller $m_\nu$, increase the absolute size of the
stop--chargino loop contribution to $b \to s \gamma$ decays. This has to be
compensated by increasing $m_0$ and/or $M_{1/2}$. The $b \rightarrow s \gamma$
constraint is therefore relatively more important in our scenario than in
mSUGRA, especially if $Y_N$ is sizable.

Note also that the region excluded because it does not permit radiative
symmetry breaking has a pronounced slope even for the larger neutrino mass,
i.e. smaller coupling $Y_N$. This shows that $m^2_{H_u}(M_{\rm SUSY})$ has
significant dependence on $m_0$, as remarked earlier, i.e. there is no
focusing behavior of this parameter. As expected from our discussion of
Fig.~\ref{fig:mnu}, this upper bound on $m_0$ becomes stronger when $Y_N$ is
increased, i.e. when $m_\nu$ is decreased. In a strip close to this excluded
region we nevertheless expect the lightest neutralino to have a large, perhaps
dominant, higgsino component; this region will therefore have a somewhat
similar phenomenology as the ``focus point'' region in mSUGRA \cite{focus},
especially as far as Dark Matter is concerned.

\setcounter{footnote}{0}

Finally, in the narrow green strips the constraint (\ref{wmap}) on the Dark
Matter relic density is satisfied; these strips would obviously look broader
if we had indicated the $2 \sigma$ allowed region, as more commonly done. The
overlap between the DM-- and $g_\mu-$allowed regions is colored in black. 

In Fig.~\ref{fig:tanb40} we find two such regions. At small $m_0$ $\tilde
\chi_1^0$ is bino--like, and achieves a sufficiently small relic density
through co--annihilation with $\tilde \tau_1$. For small $Y_N$ (left frame)
this region is strongly constrained by the bound on $b \to s \gamma$ decays.
We saw in Fig.~\ref{fig:stopstau} that increasing $Y_N$ reduces the $\tilde
\tau$ masses, making it possible to find scenarios with $m_{\tilde \tau_1}
\simeq m_{\tilde \chi_1^0}$ even if $M_{1/2}$ is large.

We just saw that for values of $m_0$ not far below the upper bound imposed by
electroweak symmetry breaking, $\tilde \chi_1^0$ has a sizable higgsino
component. For some range of parameters it achieves the correct relic density
mostly through annihilation into channels involving weak gauge bosons. As in
mSUGRA, this second DM--allowed region extends to very large values of $m_0$
and $M_{1/2}$, with $\tilde \chi_1^0$ becoming increasingly higgsino--like
(and therefore co--annihilation with $\tilde \chi_2^0$ and $\tilde \chi_1^\pm$
becoming increasingly important \cite{higgsino}.)

\begin{figure}[!tp]
\includegraphics[width=8.2cm,height=8.2cm,angle=270]{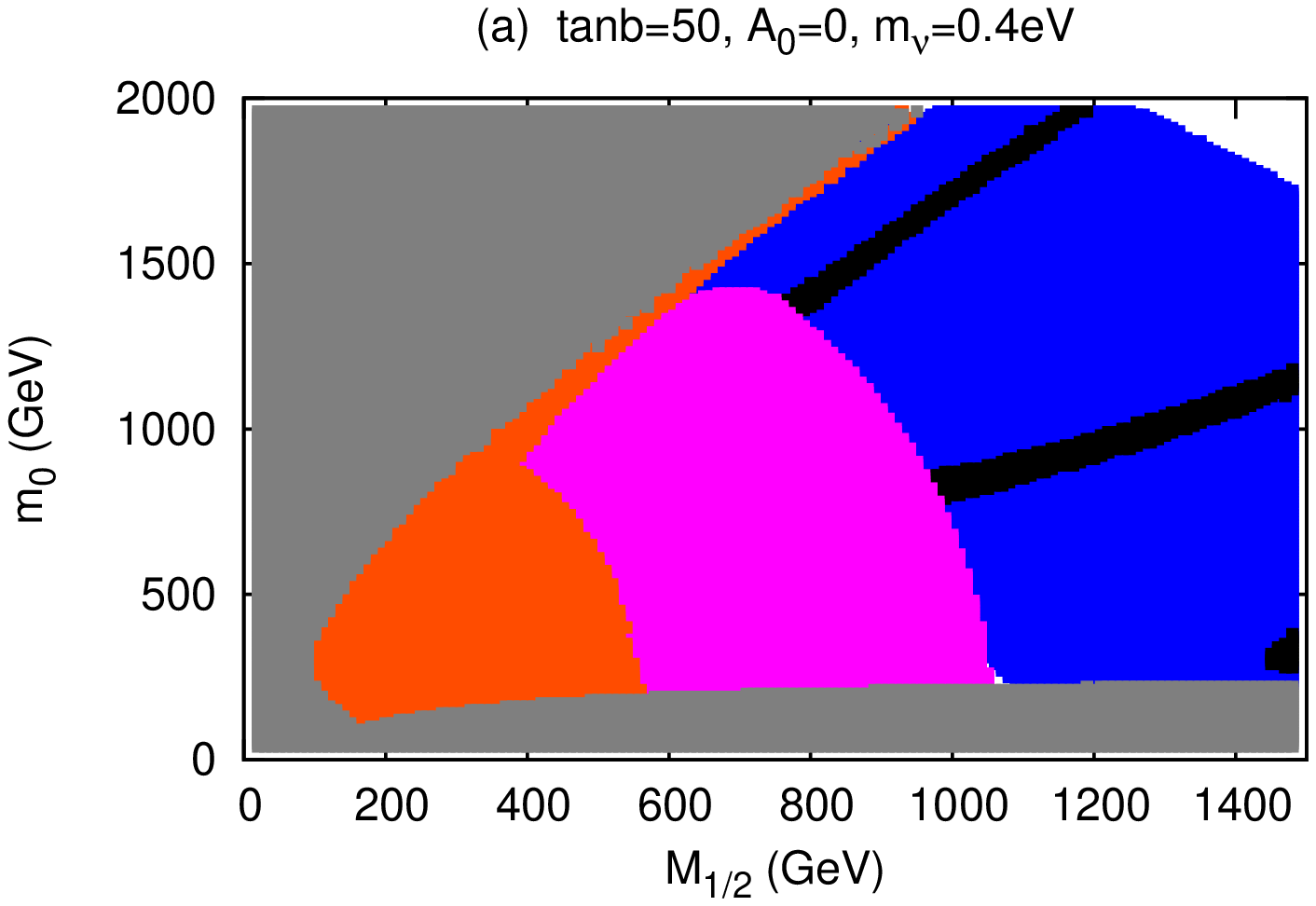}
\includegraphics[width=8.2cm,height=8.2cm,angle=270]{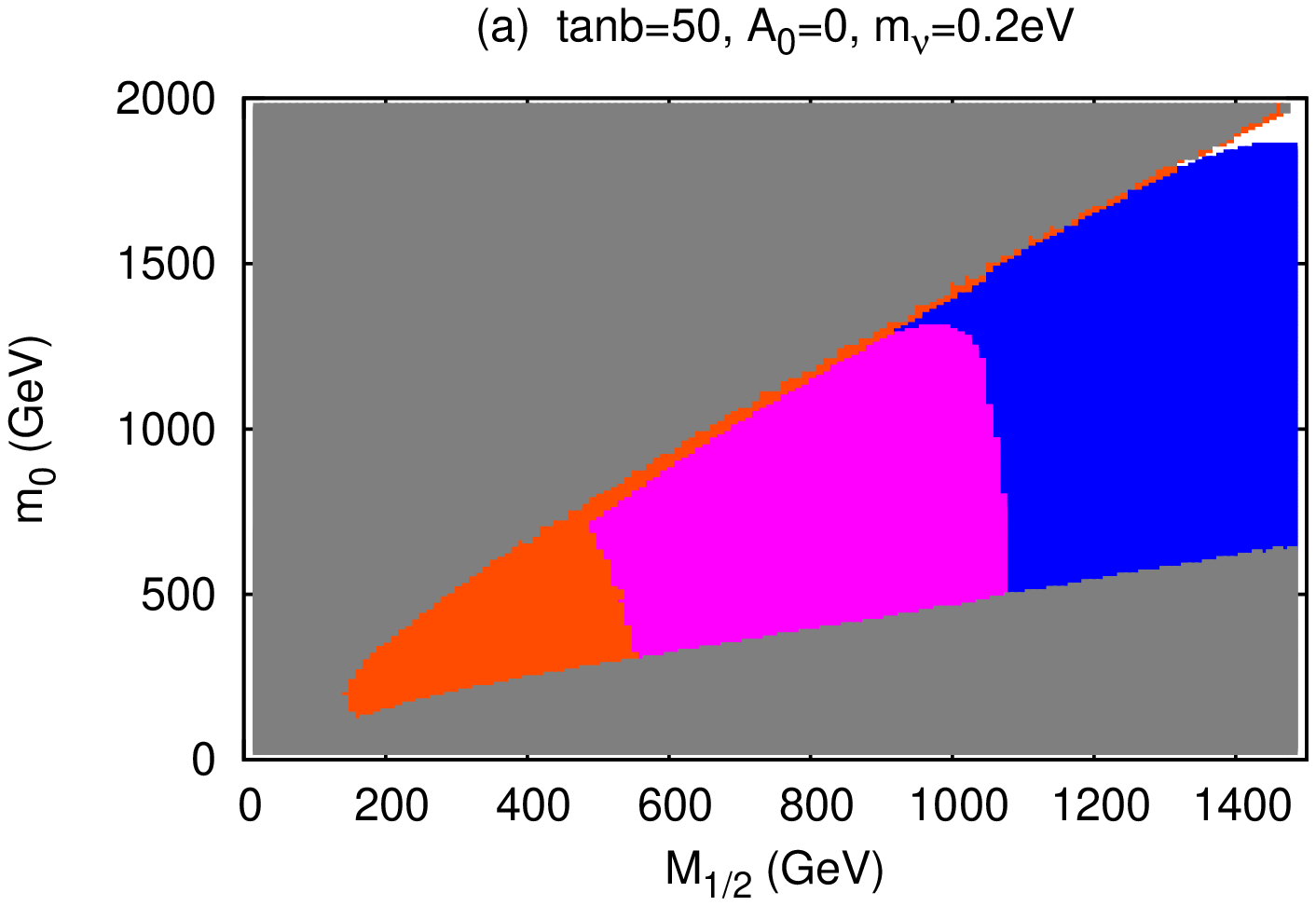}
\caption{Constraints on the $(m_0, M_{1/2})$ plane of our model. Parameter
  values and color code are the same as in Fig.~\ref{fig:tanb40}, except that
  $\tan\beta$ has been increased to 50.} 
    \label{fig:tanb50}
\end{figure}

As in mSUGRA, $\tan\beta = 40$ is not large enough to allow $m_A \simeq 2
m_{\tilde \chi_1^0}$ if $\mu > 0$. Fig.~\ref{fig:tanb50} shows that this
``$A-$pole'' region becomes accessible for $\tan\beta = 50$. Sufficiently
small values of $m_A$ are only possible if the soft breaking mass $m^2_{H_d}$
of the second Higgs boson also becomes negative (and large) at the weak scale.
We saw in the discussion of Fig.~\ref{fig:mx} that decreasing $m_\nu$ will
increase $m^2_{H_u}(M_{\rm SUSY})$. Indeed, in Fig.~\ref{fig:tanb50} we find a
well--defined $A-$funnel only for $m_\nu = 0.4$ eV (left frame). 

\begin{figure}[h!]
\begin{center}
\includegraphics[width=8.2cm,height=8.2cm,angle=270]{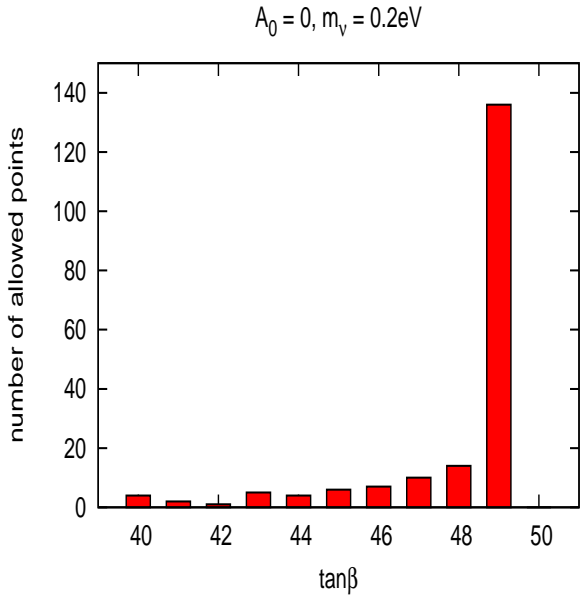}
\caption{Number of allowed points (by all constraints) for 1TeV $< m_0 <
  1.5$TeV; 1.1TeV $< M_{1/2} <1.4$TeV, with the grid $25$GeV. The allowed
region is very large when $\tan\beta = 49$.} 
    \label{fig:tanb}
\end{center}
\end{figure}

If we instead take $m_\nu = 0.2$ eV (right frame), we find that the $\tilde
\chi_1^0$ relic density becomes too low in the entire allowed region of the
$(m_0, \, M_{1/2})$ plane we scanned. One reason is that increasing $Y_N$
reduces $\mu(M_{\rm SUSY})$, as discussed above. This increases the coupling
of the LSP to neutral Higgs bosons, in particular to $A$. Since for $\tan\beta
= 50$ the $b$ and $\tau$ Yukawa couplings are quite sizable, virtual $A$
exchange diagrams become large, even though $2 m_{\tilde \chi_1^0}$ is
somewhat below $m_A$. Increasing $m_0$ increases $m_A$, but at the same time
decreases $\mu$ even further, and therefore does not allow to achieve a DM
relic density above the lower bound in the range (\ref{wmap}). Moreover,
recall that reducing $m_\nu$ also reduces the $\tilde \tau$ masses. In
addition, the very large value of $\tan \beta$ considered in this figure leads
to large $\tilde \tau_L \tilde \tau_R$ mixing, which allows $\tilde \chi_1^0
\tilde \chi_1^0 \to \tau^+ \tau^-$ annihilation through $\tilde \tau$ exchange
even if the initial state is in an $S-$wave \cite{sfermix}. Finally, for
$M_{1/2}$ close to its lower bound, $\tilde \chi_1^0 \tilde \tau_1$
co--annihilation again becomes important. Note that this indicates that the 
DM-allowed region may be quite large for some $\tan\beta$ between 40 and 50, 
and $m_\nu = 0.2$ eV. Indeed, Fig.~\ref{fig:tanb} shows that for $\tan\beta =
49$, about 50\% of the points we scanned that satisfy the other constrains are
also compatible with the DM constraint.

\begin{figure}[!h]
\includegraphics[width=8.2cm,height=8.2cm,angle=270]{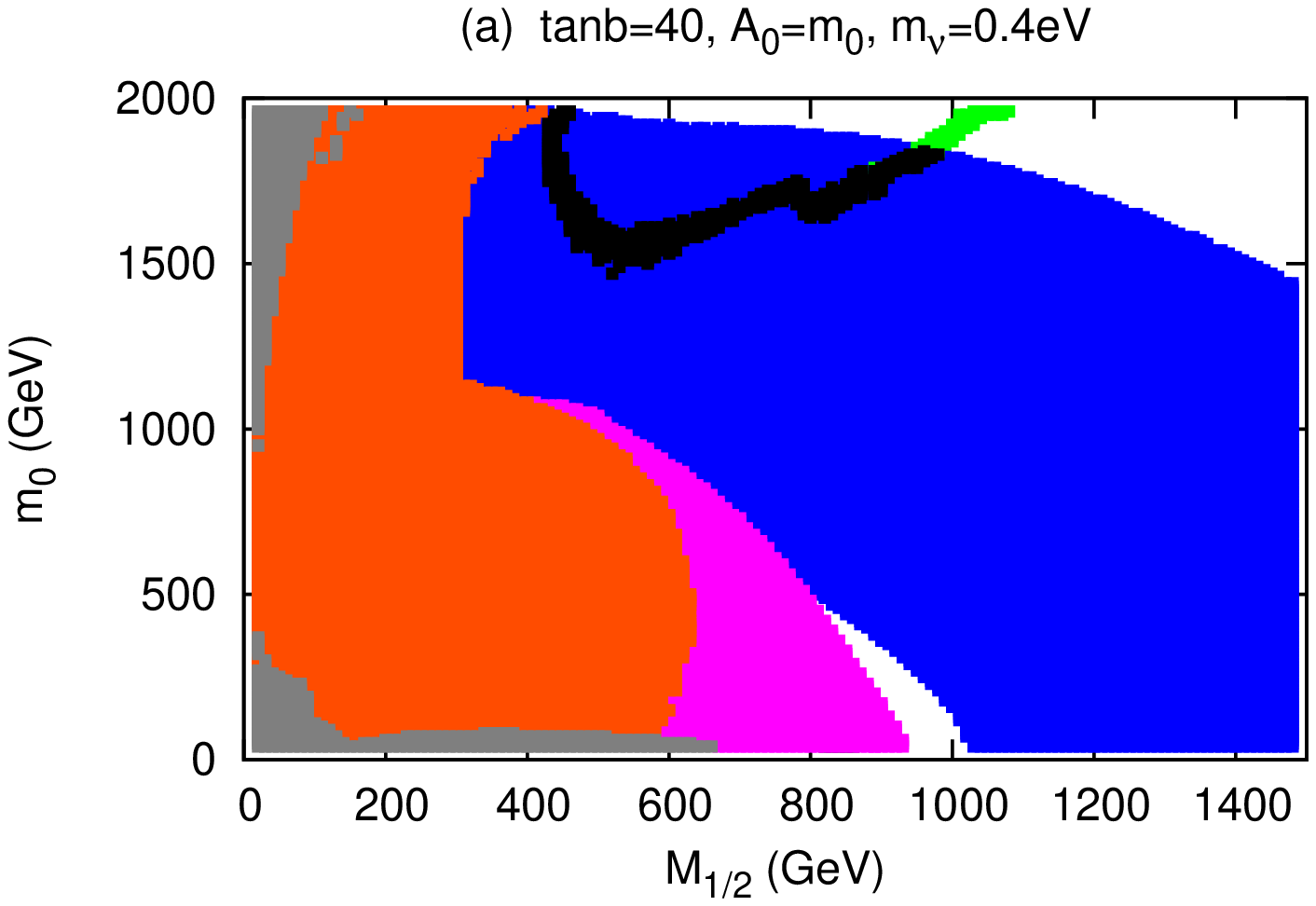}
\includegraphics[width=8.2cm,height=8.2cm,angle=270]{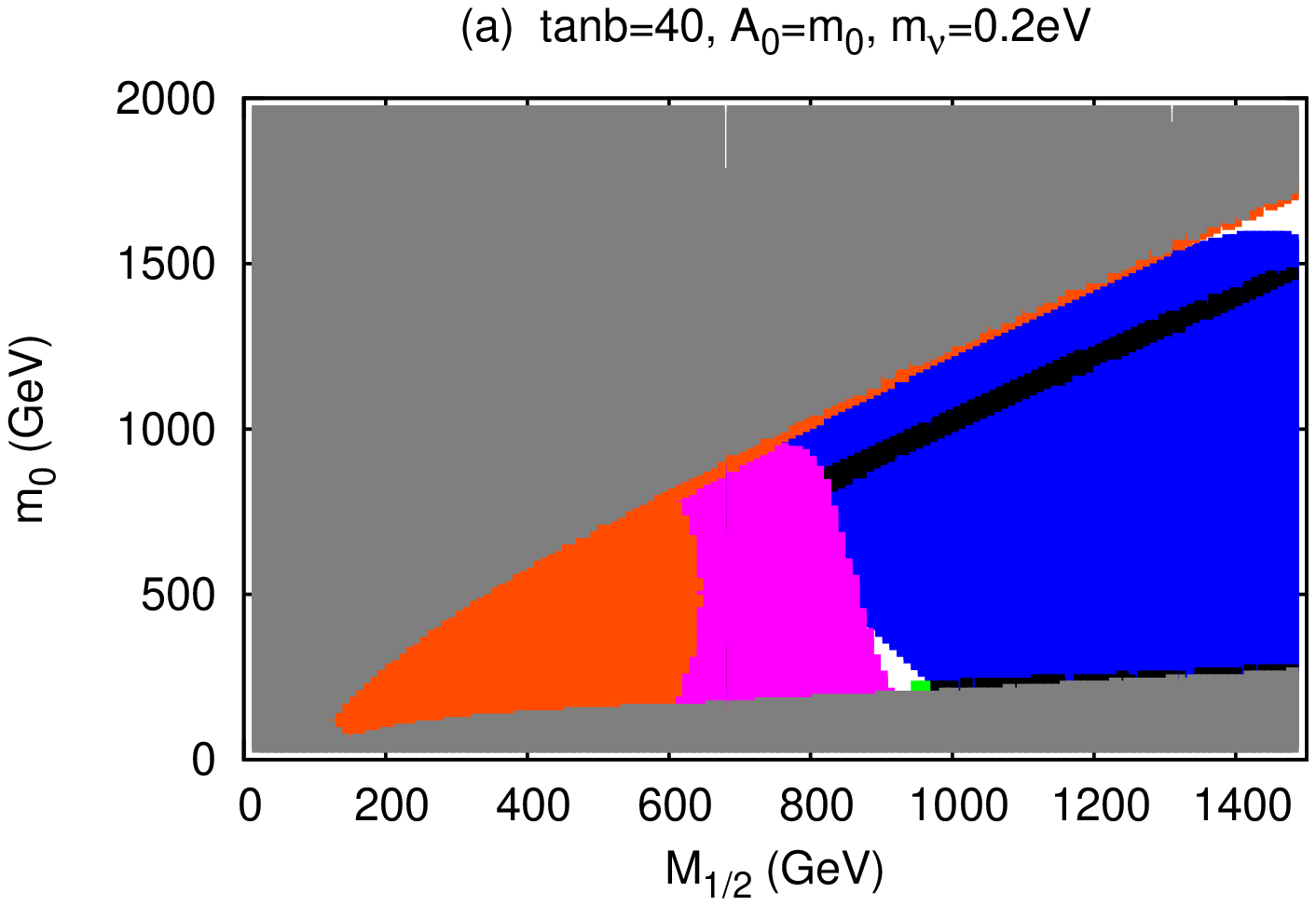}\\
\includegraphics[width=8.2cm,height=8.2cm,angle=270]{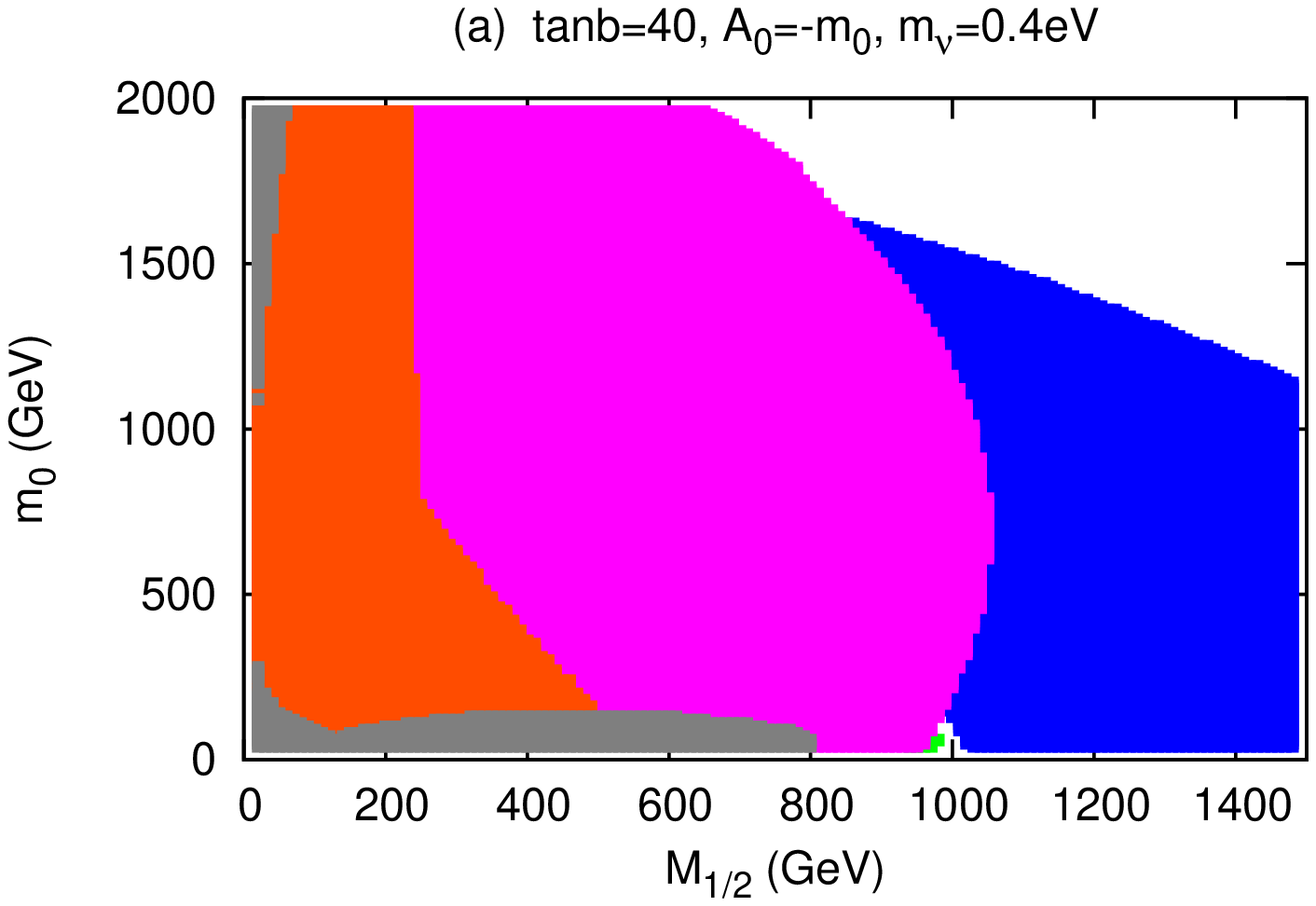}
\includegraphics[width=8.2cm,height=8.2cm,angle=270]{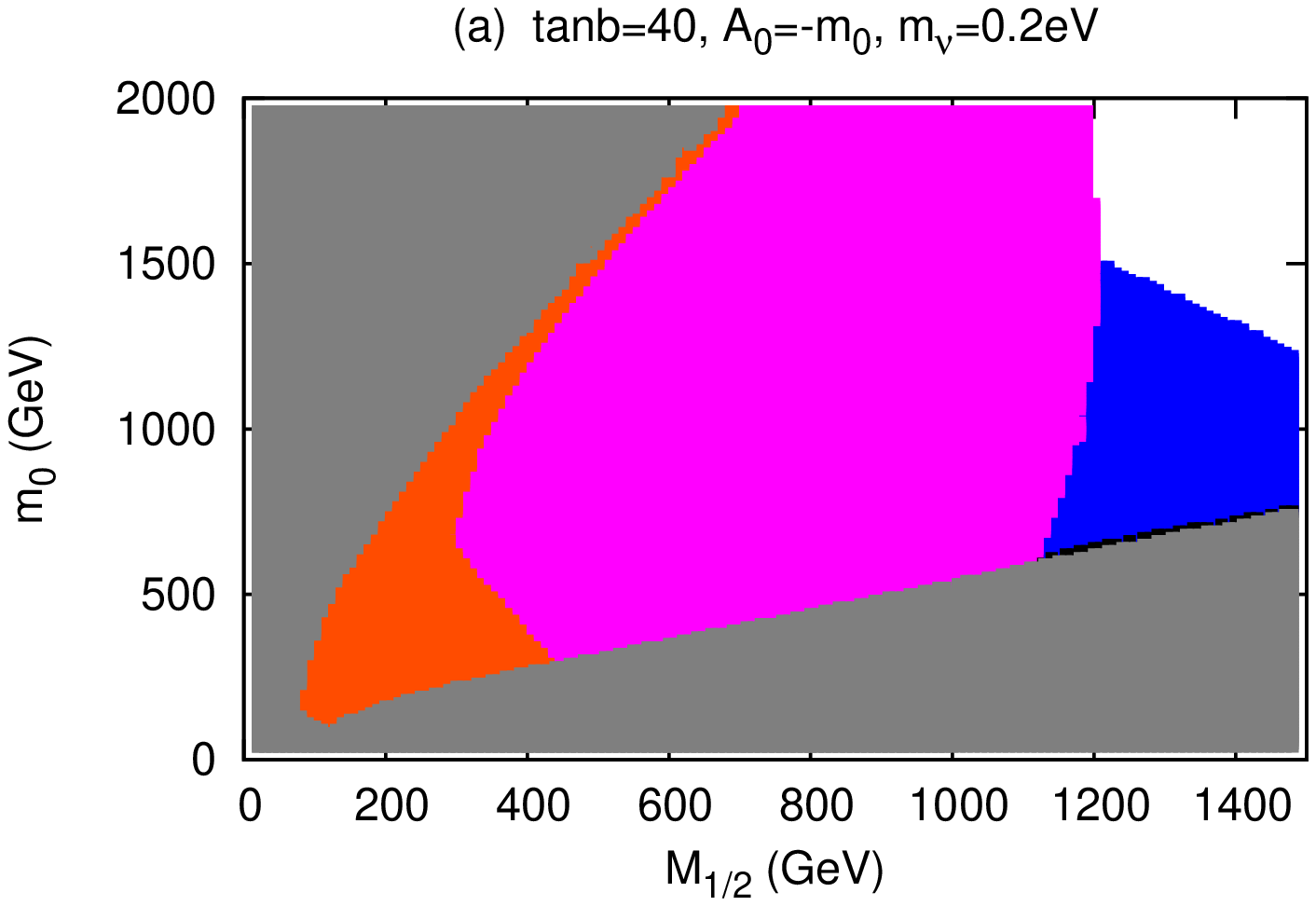}
  \caption{Constraints on the $(m_0, M_{1/2})$ plane of our model. Parameter
  values and color code are the same as in Fig.~\ref{fig:tanb40}, except that
    we now take $A_0 = m_0$ ($A_0 = -m_0$) in the top (bottom) row.} 
    \label{fig:tanb40_a0m0}
\end{figure}

In Fig.~\ref{fig:tanb40_a0m0} we explore the effect of taking a non--zero
value of $A_0$. We see that the value of $A_0$ can have quite a dramatic
effect on the region excluded because it does not allow electroweak symmetry
breaking. This can be understood as follows. By dimensional arguments and the
fact that scalar masses always appear as squares in the RGE, the soft breaking
mass of the up--type Higgs boson at the weak scale can be written as
\begin{equation} \label{mhu}
m^2_{H_u}(M_{\rm SUSY}) = a m_0^2 + b M^2_{1/2} + c A_0^2 + d M_{1/2} A_0\,.
\end{equation}
The values of the coefficients $a,b,c,d$ depend on the dimensionless couplings
in the theory, as well as (logarithmically) on $M_{\rm SUSY}$. In our model,
$a$ and $d$ are positive while $b$ and $c$ are negative. Hence increasing
$m_0$ makes EWSB more difficult, while increasing $M_{1/2}$ makes it easier if
$M_{1/2} > |A_0|$. This explains the qualitative feature of the regions
excluded by the EWSB constraint in Figs.~\ref{fig:tanb40} and
\ref{fig:tanb50}.

On the other hand, if $|A_0| \gg M_{1/2}$, increasing the absolute value of
$A_0$ also aids EWSB independent of its sign. This explains why the region
excluded by the EWSB constraint becomes much smaller in the two left frames of
Fig.~\ref{fig:tanb40_a0m0}. Fig.~\ref{fig:tanb40} shows that, for the given
small value of $Y_N$, the EWSB constraint only excludes scenarios with $m_0 >
M_{1/2}$ even if $A_0 = 0$. In the critical region $|A_0| = m_0$ is thus
always sufficiently larger than $M_{1/2}$. Finally, for given absolute value
of $A_0$, EWSB will be easier for negative than for positive $A_0$. This
explains why the EWSB excluded region is significantly larger in the
upper--right frame of Fig.~\ref{fig:tanb40_a0m0} than in the lower--right
frame. Note also that a sizable $Y_N$ decreases the absolute size of $c$,
since $Y_N$ reduces $|A_t|$ for $Q > M_C$, see Eq.(A.34)

A nonvanishing $A_0$ also changes the regions allowed by the other
constraints. In particular, $A_0 < 0$ increase $\tilde t_L - \tilde t_R$
mixing. This has two effects. On the one hand, it increases the radiatively
corrected mass of the lightest CP--even Higgs boson, thereby reducing the size
of the red regions in Fig.~\ref{fig:tanb40_a0m0}. On the other hand, it
increases the $\tilde t \tilde \chi^\pm$ contributions to radiative $b
\rightarrow s \gamma$ decays, increasing the size of the pink regions. This
latter effect completely removes the DM--allowed region close to the
EWSB--forbidden region, where $\tilde \chi_1^0$ has sizable higgsino
component. As a result, for $A_0 = -m_0$, only the small $\tilde \tau$
co--annihilation region survives. On the other hand, for $A_0 = m_0$ we again
find sizable DM--allowed regions at large $m_0$; the structure in this (black)
region at $M_{1/2} \simeq 800$ GeV in the top--left frame is due to the
opening of the $\tilde \chi_1^0 \tilde \chi_1^0 \rightarrow t \bar t$ channel.

\section{Summary and Conclusions}

Supersymmetric $SO(10)$ GUTs have become attractive extensions of the SM,
especially since the observation of the nonzero neutrino mass. However, there
is a small discrepancy between the order of the expected mass of the right
handed neutrino in the seesaw mechanism and the GUT scale; it can be explained
in a natural way if one postulates intermediate scales where the gauge
symmetry is larger than that of the SM, but smaller than $SO(10)$.

Therefore, in this work we chose a model \cite{aulakh} which gives us
intermediate symmetry breaking scale(s), and analyzed how this affects the low
energy phenomenology. We found that the relation between weak--scale and
GUT--scale parameters is quite different in this model than in the widely
considered mSUGRA scenario. Perhaps more importantly, ratios of different
weak--scale masses also differ from mSUGRA. In particular, the slepton to
electroweak gaugino mass ratios are higher than in mSUGRA. As a result,
co--annihilation is only possible with the lighter $\tilde \tau$ eigenstate,
and only at large $\tan\beta$ and/or large Yukawa coupling $Y_N$ of the SM
singlet neutrinos; the latter corresponds to small values for the light
neutrino masses.

Radiative electroweak symmetry breaking also is more difficult in this model
than in mSUGRA. This makes it easier to find Dark Matter allowed solutions
where the lightest neutralino has a significant higgsino component. As in
mSUGRA, the location of this region strongly depends on $A_0$; in addition, we
find a strong dependence on $Y_N$, i.e. on the light neutrino mass. We also
found that for very large $\tan\beta$ and large $Y_N$ most of the $(m_0,
M_{1/2})$ plane leads to too small a $\tilde \chi_1^0$ relic density. As a
corollary, there exist combinations of $Y_N$ and $\tan\beta$ where
$\Omega_{\rm DM} h^2$ has weak dependence on $m_0$ and $M_{1/2}$; however, in
this case it depends strongly on $\tan\beta$ and $Y_N$. Finally, as in mSUGRA
the $A-$pole region only exists at large $\tan\beta$; it disappears for large
values of $Y_N$.

We would like to point it out that, even though our analysis is done for a
specific model, many of our results should remain qualitatively correct for
other $SO(10)$ GUT scenarios, as long as the seesaw mechanism at an
intermediate scale plays a role. In particular, the relation between the
right--handed stau mass and the Majorana Yukawa coupling $Y_N$, which largely
determines the behavior of the co--annihilation region, does not depend on
the details of either the symmetry breaking chain or the seesaw structure. Any
partial unification above the see--saw scale also implies that $Y_N$ will
affect other sfermion masses, and hence the conditions for radiative symmetry
breaking. 

In summary, the model we considered relates several phenomena, and can hence
be probed through a large variety of measurements, from proton decay (which
imposes limits on the GUT scale) over neutrino masses and Dark Matter physics
to collider physics. We intend to investigate characteristic features of this
scenario at the LHC in a future publication.

\subsection*{Acknowledgments}
JMK thanks to C.~S.~Aulakh and M.~Kakizaki for useful discussions. This work
was partially supported by the Marie Curie Training Research Network
``UniverseNet'' under contract no.  MRTN-CT-2006-035863, as well as by the
European Network of Theoretical Astroparticle Physics ENTApP ILIAS/N6 under
contract no. RII3-CT-2004-506222. JMK was partially supported by the
Bonn--Cologne Graduate School of Physics and Astronomy. We thank the KIAS
school of physics for hospitality while part of this work was done.

\appendix
\section{Renormalization Group Equations}
\renewcommand{\theequation}{A.\arabic{equation}}
\setcounter{equation}{0}

In this section we list all relevant one--loop renormalization group equations
explicitly. Our calculations are based on the general expressions of
ref.\cite{mv}. We divide the entire energy range between the SUSY and GUT
scales into five regions, with different particles participating in the RGE
and different symmetry groups:
\begin{itemize}
  \item Region I  ($M_{\rm SUSY} < Q < M_2$) : $SU(3)_C \times SU(2)_L \times
  U(1)_Y$  
  \item Region II ($M_2 < Q < M_R$) : $SU(3)_C \times SU(2)_L \times U(1)_Y$
  \item Region III  ($M_R < Q < M_1$) : $SU(3)_C \times U(1)_{B-L} \times
    SU(2)_L \times SU(2)_R$
  \item Region IV   ($M_1 < Q < M_C$) : $SU(3)_C \times U(1)_{B-L} \times
    SU(2)_L \times SU(2)_R$
  \item Region V  ($M_C < Q < M_X$) : $SU(4)_C \times SU(2)_L \times SU(2)_R
    \times D$  
\end{itemize}

In the following Subsections we discuss the running of the supersymmetric
parameters (gauge couplings and parameters of the superpotential) and of the
soft breaking parameters, respectively.

\subsection{Superpotential Parameters}

We begin with the parameters that preserve supersymmetry. The running of the
gauge couplings is described by
\begin{eqnarray}\label{eq:rge_cc}
\frac{d}{dt} g_a = \frac{1}{16\pi^2}\beta_{g_a} g_a^3 \ \ \ {\rm with} \ \
\beta_{g_a} =  \sum_R S(R) - 3 C_a(G) \,.
\end{eqnarray}
Here $t = \ln (Q / Q_0), \ a$ labels the factor group, $R$ the representation
of the matter and Higgs superfields under this group, $C_a$ is the
quadratic Casimir of this group, and the Dynkin index $S(R)$ is defined by
${\rm Tr}(t_A t_B) = S(R) \delta_{AB}$, $t_{A,B}$ being matrix representations
of the gauge group.

Our notation for a generic superpotential is  
\begin{eqnarray} \label{W}
W = \frac{1}{6}Y^{ijk}\Phi_i \Phi_j \Phi_k + \frac{1}{2}\mu^{ij}\Phi_i
  \Phi_j \,.
\end{eqnarray}
The running of the parameters appearing in (\ref{W}) is  given by
\begin{eqnarray}\label{eq:rge_yukawa}
\frac{d}{dt} Y^{ijk} &=& Y^{ijp}\frac{1}{16\pi^2}\gamma_p^k +
  (k \leftrightarrow i) + (k \leftrightarrow j) 
\nonumber \\
\frac{d}{dt} \mu^{ij} &=& \mu^{ip}\frac{1}{16\pi^2}\gamma_p^j + (j
  \leftrightarrow i) \,,
\end{eqnarray}
where summation over repeated indices is understood. The anomalous dimensions
$\gamma_i^j$ are given by
\begin{eqnarray}\label{eq:rge_anom}
\gamma_i^j = \frac{1}{2}Y_{ipq}Y^{jpq} - 2\delta_i^j g_\alpha^2 C_\alpha(i).
\end{eqnarray}

In our case, the superpotential below $M_X$ has been given in
Eq.(\ref{Y_MSSM}) for region I, in Eq.(\ref{Y_gen}) for region II, in
Eq.(\ref{Y_3122}) for regions III and IV, and in Eq.(\ref{Y_422}) for region
V. Recall that we take $Y_2 = Y_{q,2} = Y_{l,2} = Y_{u,2} = Y_{d,2} = Y_{e,2}
= 0$; for the sake of simplicity we therefore suppress the superscript 1 on
the Yukawa couplings in the following. These couplings are $3\times 3$
matrices in generation space. We will write the RGE for general matrices,
although we only kept third generation couplings in our numerical analysis. We
use the general notation
\begin{eqnarray} 
\frac{d}{dt}Y_f &=& \frac{1}{16\pi^2}\beta_{Y_f} \,;
\nonumber \\
\frac{d}{dt} \mu &=& \frac{1}{16\pi^2}\beta_\mu\,,
\end{eqnarray}
where $f$ stands for any matter fermion.  In the following we list these as
well as the gauge beta--functions in the five different energy regions.

\subsubsection{Region I}

The coefficients of the gauge beta functions are
\begin{equation}
\beta_{g_a} = (33/5, 1, -3) \ \ {\rm for} \ a = (1_Y, 2_L, 3_C)\,,
\end{equation}
where we have used GUT normalization for the $U(1)_Y$ factor. The
corresponding coefficients for the MSSM Yukawa couplings are
\begin{eqnarray} \label{beta_yuk}
\beta_{Y_u} &=& Y_u(\gamma_U^U + \gamma_{H_u}^{H_u}) + \gamma_Q^Q Y_u \,;
\nonumber \\
\beta_{Y_d} &=& Y_d(\gamma_D^D + \gamma_{H_d}^{H_d}) + \gamma_Q^Q Y_d \,;
\nonumber \\
\beta_{Y_e} &=& Y_e(\gamma_E^E + \gamma_{H_d}^{H_d}) + \gamma_L^L Y_e \,;
\nonumber \\
\beta_\mu &=& \mu(\gamma_{H_d}^{H_d} + \gamma_{H_u}^{H_u}) \,,
\end{eqnarray}
where
\begin{eqnarray} \label{gamma_1}
\gamma_E^E &=& 2Y_e^{\dagger}Y_e - \frac{6}{5}g_1^2 \,;
\nonumber \\
\gamma_L^L &=& Y_e Y_e^{\dagger} -
      \frac{3}{10}g_1^2 - \frac{3}{2}g_2^2  \,;
\nonumber \\
\gamma_Q^Q &=& Y_d Y_d^{\dagger} + Y_u Y_u^{\dagger} - \frac{1}{30}g_1^2 -
      \frac{3}{2}g_2^2 - \frac{8}{3}g_3^2  \,;
\nonumber \\
\gamma_U^U &=& 2Y_u^{\dagger} Y_u - \frac{8}{15}g_1^2 - \frac{8}{3}g_3^2 \,;
\nonumber \\
\gamma_D^D &=& 2Y_d^{\dagger} Y_d - \frac{2}{15}g_1^2 - \frac{8}{3}g_3^2 \,;
\nonumber \\
\gamma_{H_d}^{H_d} &=& {\rm tr}(3Y_d Y_d^{\dagger} + Y_e Y_e^{\dagger}) -
      \frac{3}{10}g_1^2 - \frac{3}{2}g_2^2 \,;
\nonumber \\
\gamma_{H_u}^{H_u} &=& 3{\rm tr}Y_u Y_u^{\dagger} -
      \frac{3}{10}g_1^2 - \frac{3}{2}g_2^2 \,.
    \end{eqnarray}

\subsubsection{Region II}

\begin{eqnarray}
\beta_{g_a} = (12, 2, -3) \ \ {\rm for} \ \ a = (1_Y, 2_L, 3_C)\,.
\end{eqnarray}
The Yukawa coupling beta functions of the MSSM matter fields have the same
form as in Region I, but we need to introduce an RGE for
$Y_N$:
\begin{equation} \label{beta_Y_N}
\beta_{Y_N} = Y_N (\gamma_E^E + \gamma_{\bar \delta}^{\bar \delta} ) +
\gamma_E^E Y_N\,.
\end{equation}
Except for $\gamma_E^E$ the anomalous dimensions of the MSSM matter fields
also remain form invariant, and we have to introduce an anomalous dimension
for $\bar \delta^{--}$:
\begin{eqnarray} \label{gamma_e_e}
\gamma_E^E &=& 2Y_e^{\dagger}Y_e + Y_N^\dagger Y_N -
\frac{6}{5}g_1^2 \,; \nonumber \\
\gamma_{\bar \delta}^{\bar \delta} &=& \frac{1}{2} {\rm tr} (Y_N^\dagger Y_N)
- \frac {24} {5} g_1^2\, .
\end{eqnarray}
Recall that we are now dealing with the couplings $Y_{f,1} \ (f = u,d,e)$,
which are related to the MSSM couplings $Y_f$ via
Eqs.(\ref{match})--(\ref{phi_d}).

\subsubsection{Region III}

\begin{eqnarray}
\beta_{g_a} = (15, 2, 6, -3) \ \ {\rm for} \ a = (1_{B-L}, 2_L, 2_R, 3_C)\,,
\end{eqnarray}
where we have again used GUT normalization for the $U(1)$ coupling. The
effective coefficient $48/5$ for the running $U(1)_Y$ coupling listed in
Table~2 is $\frac{3}{5} \cdot 6 + \frac{2}{5} \cdot 15$, which follows from
the matching condition $g_Y^{-2} = \frac{3}{5} g_R^{-2} + \frac {2}{5}
g_{B-L}^{-2}$. 

Since the underlying symmetry group is enhanced, and the matter superfields
form multiplets correspondingly, their anomalous dimensions receive
contributions from the heavy gauge bosons that become active in this energy
range. We switch to the notation of Eq.(\ref{Y_3122}), i.e. we introduce
$Q^c$ instead of $U^c$ and $D^c$, while $E^c$ and $N^c$ are united in $L^c$
and $H_u$ and $H_d$ are united in $\Phi$. The number of independent Yukawa
couplings is thus reduced to three:
\begin{eqnarray} \label{beta_yuk2}
\beta_{Y_q} &=& Y_q ( \gamma_{Q^c}^{Q^c} + \gamma_\Phi^\Phi ) + \gamma_Q^Q Y_q
\,; \nonumber \\
\beta_{Y_l} &=& Y_l(\gamma_{L^c}^{N^c} + \gamma_\Phi^\Phi) + \gamma_L^L
Y_l \,; 
\nonumber \\
\beta_{Y_N} &=& Y_N(\gamma_N^N + \gamma_{\bar{\delta}}^{\bar{\delta}}) +
      \gamma_N^N Y_N \,.
\end{eqnarray}
The relevant anomalous dimensions read:
\begin{eqnarray} \label{gamma_3}
\gamma_{L^c}^{L^c} &=& 2 Y_l^{\dagger}Y_l + \frac{3}{2}Y_N^{\dagger}Y_N -
      \frac{3}{2}g_R^2 - \frac{3}{4}g_{B-L}^2 \,;
\nonumber \\
\gamma_L^L &=& 2 Y_l Y_l^{\dagger} -
      \frac{3}{2}g_2^2 - \frac{3}{4}g_{B-L}^2  \,;
\nonumber \\
\gamma_Q^Q &=& 2 Y_q Y_q^{\dagger}  - \frac{3}{2}g_2^2 -
      \frac{1}{12}g_{B-L}^2 - \frac{8}{3}g_3^2  \,;
\nonumber \\
\gamma_{Q^c}^{Q^c} &=& 2Y_q^{\dagger} Y_q - \frac{3}{2}g_R^2 -
      \frac{1}{12}g_{B-L}^2 - \frac{8}{3}g_3^2  \,;
\nonumber \\
\gamma_{\Phi}^{\Phi} &=& {\rm tr}(3Y_q Y_q^{\dagger} + Y_l Y_l^{\dagger}) -
      \frac{3}{2}g_R^2 - \frac{3}{2}g_2^2 \,;
\nonumber \\
\gamma_{\bar{\delta}}^{\bar{\delta}} &=& \frac{1}{2}{\rm tr}(Y_N Y_N^{\dagger})
      - 4g_R^2 - 3g_{B-L}^2       \,.
\end{eqnarray} 
Here we have continued to use $g_2$ for the $SU(2)_L$ coupling, and denoted
the $SU(2)_R$ coupling with $g_R$. Eqs.(\ref{gamma_3}) are consistent with
\cite{ss}, taking the appropriate normalization.

\subsubsection{Region IV}
\setcounter{footnote}{0}

\begin{eqnarray}
\beta_{g_a} = (15, 2, 6, 0) \ \ {\rm for} \ a = (1_{B-L}, 2_L, 2_R, 3_C) \,.
\end{eqnarray}
Since the new massive fields becoming active in this energy range are singlets
under $U(1)_{B-L} \times SU(2)_R \times SU(2)_L$, only the running of the
$SU(3)_C$ group changes. Moreover, the Yukawa coupling beta functions are
those of Region III. 

\subsubsection{Region V}

\begin{eqnarray}
\beta_{g_a} = (42, 42, 34) \ \ {\rm for} \ a = (2_L, 2_R, 4_C)\,.
\end{eqnarray}
Since many new fields become active at $Q \geq M_C$, all gauge
$\beta-$functions increase quite dramatically. In GUT normalization, $g_{B-L}
= g_3 = g_4$, where $g_4$ is the $SU(4)_C$ gauge coupling; this explains the
entries in the last row of Table~2, with $\frac {194}{5} = \frac{3}{5} \cdot
42 + \frac{2}{5} \cdot 34$.

No new Yukawa couplings appear in this energy range; instead, the couplings
$Y_q$ and $Y_l$ get unified into the single coupling $Y$. At the same time,
all MSSM matter superfields are now in $F$ or $F^c$ introduced in
Eq.(\ref{Y_422}); the $D$ symmetry ensures that the anomalous dimensions of
these two superfields are the same. Moreover, the $SU(2)_R$ triplet Higgs
superfield $\bar \delta$ gets embedded into the much larger representation
$\overline{\Sigma}_R$, and the $D-$partner $\overline{\Sigma}_L$ also
appears, with identical anomalous dimensions. 

Therefore Eqs.(\ref{beta_yuk2}) change to
\begin{eqnarray} \label{beta_yuk3}
\beta_{Y} &=& Y ( \gamma_F^F + \gamma_\Phi^\Phi ) + \gamma_F^F Y
\,; \nonumber \\
\beta_{Y_N} &=& Y_N(\gamma_F^F +
\gamma_{\overline{\Sigma}_R}^{\overline{\Sigma}_R}) +  \gamma_F^F Y_N \,.
\end{eqnarray}
The anomalous dimensions appearing in Eqs.(\ref{beta_yuk3}) are:
\begin{eqnarray}
\gamma_F^F &=& 2Y^{\dagger}Y + \frac{15}{4}Y_N^{\dagger}Y_N -
      \frac{3}{2}g_R^2 - \frac{15}{4}g_4^2 \,;
\nonumber \\
\gamma_{\Phi}^{\Phi} &=& {\rm tr}(4Y Y^{\dagger})  -
      \frac{3}{2}g_R^2 - \frac{3}{2}g_2^2\,;
\nonumber \\
\gamma_{\overline{\Sigma}_R}^{\overline{\Sigma}_R} &=& \frac{1}{2}{\rm tr}(Y_N
      Y_N^{\dagger}) - 4g_R^2 - 9g_4^2   \,.
\end{eqnarray} 

\subsection{Soft SUSY-breaking Parameters}

We write the part of the Lagrangian that softly breaks supersymmetry as
\begin{eqnarray} \label{lsoft}
\mathcal{L}_{SB} = -\frac{1}{6}h^{ijk}\phi_i \phi_j \phi_k 
- \frac{1}{2}b^{ij}\phi_i \phi_j - \frac{1}{2}(m^2)_i^j\phi^{*i} \phi_j 
- \frac{1}{2}M_a \lambda_a \lambda_a +  h.c. 
\end{eqnarray}
We assume universal boundary conditions,
\begin{eqnarray} \label{boundary}
h^{ijk} &=& Y^{ijk} A_0\,; \nonumber \\
(m^2)_i^j &=& m_0^2 \delta_i^j \,; \nonumber \\
M_a &=& M_{1/2} \ \forall a \,,
\end{eqnarray}
which hold at scale $Q = M_X$. Here $Y^{ijk}$ are the superpotential couplings
introduced in Eq.(\ref{W}).
 
The $\beta-$functions of the soft breaking parameters are defined by
\begin{eqnarray} \label{betasoft}
\frac{d}{dt}h^{ijk} &=& \frac{1}{16\pi^2}\beta_h^{ijk} \,; \nonumber \\
\frac{d}{dt}b^{ij} &=& \frac{1}{16\pi^2}\beta_b^{ij} \,; \nonumber \\
\frac{d}{dt}(m^2)_i^j &=& \frac{1}{16\pi^2}{\beta_{m^2}}_i^j \,; \nonumber \\
\frac{d}{dt} M_a &=& \frac {1} {16 \pi^2} 2 g_a^2 M_a^2 \beta_{g_a}\,.
\end{eqnarray}
Here $\beta_{g_a}$ are the coefficients of the gauge $\beta-$functions
introduced in Eq.(\ref{eq:rge_cc}). The other $\beta-$functions appearing in
Eqs.(\ref{betasoft}) can be written as\footnote{We suppress terms that can be
  nonzero only in the presence of complete gauge singlet chiral superfields.}
\begin{eqnarray} \label{betasoft1}
\beta_h^{ijk} &=& \frac{1}{2}h^{ijl}Y_{lmn}Y^{mnk} + Y^{ijl}Y_{lmn}h^{mnk}
-  2(h^{ijk} - 2M_aY^{ijk})g_a^2C_a(k) 
+ (k \leftrightarrow i) + (k \leftrightarrow j) \,; \nonumber \\
\beta_b^{ij} &=& \frac{1}{2}b^{il}Y_{lmn}Y^{mnj} +
+ \mu^{il}Y_{lmn}h^{mnj} 
- 2(b^{ij} - 2M_a\mu^{ij})g_a^2C_a(i) + (i \leftrightarrow j) \,;\nonumber \\
{\beta_{m^2}}_i^j &=& \frac{1}{2}Y_{ipq}Y^{pqn}(m^2)_n^j +
  \frac{1}{2}Y^{jpq}Y_{pqn}(m^2)_i^n + 2Y_{ipq}Y^{jpr}(m^2)_r^q +
  h_{ipq}h^{jpq} \nonumber\\ 
&-&  8\delta_i^jM_aM^{\dagger}_ag_a^2C_a(i) + 2g_a^2({\bf
    t}_a^A)_i^j {\rm {\rm tr}}({\bf t}_a^Am^2) \,.
\end{eqnarray}
The last term in $\beta_{m^2}$ can be nonzero only for $U(1)$ group
factors. In the case at hand it is therefore either proportional to
\begin{eqnarray}\label{S_Y}
S_Y = m^2_{H_u} - m^2_{H_d} + {\rm tr}[m_Q^2 - m_L^2 - 2m_u^2 + m_d^2 + m_e^2]
\end{eqnarray}
or to
\begin{eqnarray}\label{S_{B-L}}
S_{B-L} = \frac{1}{2} (6m_{\overline{\Sigma}}^2 - 6m_{\Sigma}^2 + {\rm
  tr}[2m_Q^2 -  2m_{Q^c}^2 - 2m_L^2 + 2m_{L^c}^2 ]) \,. 
\end{eqnarray}
For better readability, in Eqs.(\ref{S_Y}) and (\ref{S_{B-L}}), as well as in
subsequent equations, we have omitted the tildes on the subscripts of the
squared scalar soft breaking masses; moreover, we use $m_{\Sigma}$ and
$m_{\overline{\Sigma}}$ for the soft mass of whatever parts of the original
$\Sigma$ and $\overline{\Sigma}$ superfields are active in a given energy
range. Note that both $S_Y$ and $S_{B-L}$ evolve homogeneously. Since the
boundary condition (\ref{boundary}) for scalar soft breaking masses implies
$S_Y = S_{B-L} = 0$ at scale $M_X$, they vanish at all scales. For
completeness we nevertheless list these contributions in the following.

We are now ready to give explicit expressions for the soft breaking
$\beta-$functions in the energy regions defined above.

\subsubsection{Region I}

Here the RGE are those of the MSSM \cite{mssmrge}:
\begin{eqnarray} \label{soft1_AB}
\beta_{h_u} &=& h_u \left[ {\rm tr}(3Y_uY_u^{\dagger}) + 5Y_uY_u^{\dagger} +
Y_d^{\dagger}Y_d - \frac{16}{3}g_3^2 - 3g_2^2 - \frac{13}{15}g_1^2 \right]
      \nonumber\\ 
&+& Y_u \left[ {\rm tr}(6h_uY_u^{\dagger}) + 4h_uY_u^{\dagger} +
      2Y_d^{\dagger}h_d +  \frac{32}{3}g_3^2M_3 + 6g_2^2M_2 +
      \frac{26}{15}g_1^2M_1 \right] \,; \nonumber \\
\beta_{h_d} &=& h_d \left[{\rm tr}(3Y_dY_d^{\dagger} + Y_eY_e^{\dagger}) +
      5 Y_d Y_d^{\dagger} + Y_u^{\dagger}Y_u - \frac{16}{3}g_3^2 - 3g_2^2 -
      \frac{7}{15}g_1^2 \right] \nonumber\\
&+&  Y_d \left[ {\rm tr}(6h_d Y_d^{\dagger} + 2h_eY_e^{\dagger}) +
      4Y_d^{\dagger}h_d + 2Y_u^{\dagger} h_u 
+ \frac{32}{3}g_3^2M_3 + 6g_2^2M_2 + \frac{14}{15}g_1^2M_1 \right]
\,: \nonumber \\
\beta_{h_e} &=& h_e \left[ {\rm tr}(3Y_dY_d^{\dagger} + Y_eY_e^{\dagger}) +
      5Y_e^{\dagger}Y_e - 3g_2^2 - \frac{9}{5}g_1^2 \right] \nonumber\\
&+& Y_e \left[ {\rm tr}(6h_dY_d^{\dagger} + 2h_eY_e^{\dagger}) +
      4Y_e^{\dagger}h_e + 6g_2^2M_2 + \frac{18}{5}g_1^2M_1 \right]
\,.
\end{eqnarray}

\begin{eqnarray} \label{soft1_B}
\beta_{B} &=& B \left[ {\rm tr}(3Y_uY_u^{\dagger} + 3Y_dY_d^{\dagger} +
      Y_eY_e^{\dagger}) - 3g_2^2 - \frac{3}{5}g_1^2 \right] \nonumber\\
&+& \mu \left[ {\rm tr}(6h_uY_u^\dagger + 6h_dY_d^\dagger + 2h_eY_e^\dagger) +
      6g_2^2M_2 + \frac{6}{5}g_1^2M_1 \right] \,.
    \end{eqnarray}
\begin{eqnarray} \label{soft1_m}
\beta_{m^2_{H_u}} &=& 6 {\rm tr}\left[ (m^2_{H_u} + m^2_Q)Y_u^{\dagger}Y_u +
      Y_u^{\dagger}m^2_uY_u + h_u^{\dagger}h_u \right] 
- 6g_2^2|M_2|^2 - \frac{6}{5}g_1^2|M_1|^2 + \frac{3}{5}g_1^2S_Y
\,; \nonumber \\
\beta_{m^2_{H_d}} &=& {\rm tr} \left[ 6(m^2_{H_d} + m^2_Q)Y_d^{\dagger}Y_d +
      6Y_d^{\dagger}m^2_dY_d + 2(m^2_{H_d} + m^2_L)Y_e^{\dagger}Y_e +
      2Y_e^{\dagger}m^2_eY_e  \right. \nonumber \\
&&\left. \hspace*{5mm} + 6h_d^{\dagger}h_d +  2h_e^{\dagger}h_e \right]  
- 6g_2^2|M_2|^2 - \frac{6}{5}g_1^2|M_1|^2 - \frac{3}{5}g_1^2S_Y\,;
\nonumber \\
\beta_{m^2_Q} &=& (m^2_Q + 2m^2_{H_u})Y_u^{\dagger}Y_u + (m^2_Q +
 2m^2_{H_d})Y_d^{\dagger}Y_d + [Y_u^{\dagger}Y_u + Y_d^{\dagger}Y_d]m^2_Q
+ 2Y_u^{\dagger}m^2_uY_u \nonumber\\ &+& 2Y_d^{\dagger}m^2_dY_d  
  + 2h_u^{\dagger}h_u + 2h_d^{\dagger}h_d 
  - \frac{32}{3}g_3^2|M_3|^2 - 6g_2^2|M_2|^2 - \frac{2}{15}g_1^2|M_1|^2 -
      \frac{1}{5}g_1^2S_Y \,;\nonumber\\
\beta_{m^2_L} &=& (m^2_L + 2m^2_{H_d})Y_e^{\dagger}Y_e +
      2Y_e^{\dagger}m^2_eY_e + Y_e^{\dagger}Y_em^2_L + 2h_e^{\dagger}h_e
- 6g_2^2|M_2|^2 - \frac{6}{5}g_1^2|M_1|^2 - \frac{3}{5}g_1^2S_Y
\,; \nonumber \\
\beta_{m^2_u} &=& (2m^2_u + 4m^2_{H_u})Y_uY_u^{\dagger} +
      4Y_um^2_QY_u^{\dagger} + 2Y_uY_u^{\dagger}m_u^2 + 4h_uh_u^{\dagger}
- \frac{32}{3}g_3^2|M_3|^2 - \frac{32}{15}g_1^2|M_1|^2
\nonumber \\ &-& \frac{4}{5}g_1^2 S_Y \,; \nonumber \\
\beta_{m^2_d} &=& (2m_d^2 + 4m^2_{H_d})Y_dY_d^{\dagger} +
4Y_dm^2_QY_d^{\dagger} + 2Y_dY_d^{\dagger}m_d^2 + 4h_dh_d^{\dagger}
- \frac{32}{3}g_3^2|M_3|^2 - \frac{8}{15}g_1^2|M_1|^2 \nonumber \\
&+&   \frac{2}{5}g_1^2S_Y \,; \nonumber \\
\beta_{m^2_e} &=& (2m_e^2 + 4m^2_{H_d})Y_eY_e^{\dagger} +
      4Y_em^2_LY_e^{\dagger} + 2Y_eY_e^{\dagger}m_e^2 + 4h_eh_e^{\dagger}
- \frac{24}{5}g_1^2|M_1|^2 + \frac{6}{5}g_1^2S_Y\,.
    \end{eqnarray}

\subsubsection{Region II}

Most expression from Region I remain form--invariant; however, the Yukawa
couplings should now be interpreted as the high--scale couplings $Y_{f,1}$
rather than as low--scale (MSSM) couplings $Y_f$. In addition, the
beta--function for the $SU(2)_L$ singlet slepton mass changes, and we have to
introduce beta--function for $m_{\overline{\Sigma}}$ as well as $h_N$:
\begin{eqnarray} \label{beta_hn}
\beta_{h_N} &=& h_N[\frac{1}{2}{\rm tr}(Y_N^{\dagger}Y_N) + 4Y_eY_e^{\dagger} 
      + 2Y_NY_N^{\dagger} - \frac{36}{5}g_1^2]
\nonumber\\
&+& Y_N[Y_N^{\dagger}h_N + 8Y_e^{\dagger} h_e + 4Y_N^{\dagger}h_N +
      \frac{72}{5}g_1^2|M_1|^2]\,.
\end{eqnarray}
\begin{eqnarray} \label{beta_B}
\beta_{B_\Sigma} &=& B_\Sigma [\frac{1}{2}{\rm tr}(Y_N Y_N^{\dagger})-
\frac{48}{5}g_1^2] + M_\Sigma [{\rm tr}0(Y_N^{\dagger}h_N) +
\frac{96}{5}g_1^2|M_1|^2]\,.
\end{eqnarray}
\begin{eqnarray} \label{beta_me}
\beta_{m^2_e} &=& (2m_e^2 + 4m^2_{H_d})Y_eY_e^{\dagger} +
      4Y_e m^2_L Y_e^{\dagger} + 2Y_eY_e^{\dagger} m_e^2 + 
      m_e^2Y_NY_N^{\dagger} + 2Y_N^{\dagger}m_e^2Y_N 
\nonumber\\
&+& Y_NY_N^{\dagger}m_e^2 + 2Y_N^{\dagger}m^2_{\overline{\Sigma}} Y_N +
(4m_e^2 + 2 m_{\overline{\Sigma}}^2) Y_N Y_N^\dagger +
4h_eh_e^{\dagger} +2 h_N h_N^\dagger 
\nonumber\\
&-& \frac{24}{5}g_1^2|M_1|^2 + \frac{6}{5}g_1^2S_Y\,; \nonumber\\
\beta_{m^2_{\overline{\Sigma}}} &=&
      {\rm tr}[\frac{1}{2}Y_N^{\dagger}Y_Nm^2_{\bar{\Sigma}} +
      2Y_N^{\dagger}m_e^2Y_N + \frac{1}{2}Y_NY_N^{\dagger}m^2_{\bar{\Sigma}}]
      + h_N h_N^\dagger - \frac{96}{5}g_1^2|M_1|^2 + \frac{12}{5}g_1^2S_Y\,.
\end{eqnarray}
$M_\Sigma$ appearing in the Eq.(\ref{beta_B}) is the supersymmetric
$\Sigma$ and $\overline{\Sigma}$ mass, which comes from a term $M_\Sigma
\Sigma \overline{\Sigma}$ in the superpotential. Note that to one--loop order
$B_\Sigma$ does not appear on the right--hand side of any other RGE, hence it
has no impact on the low--energy spectrum. We nevertheless list its RGE for
completeness; it might be relevant, e.g., for the detailed dynamics of
intermediate--scale symmetry breaking, which we here merely parameterize
through the vev $\sigma$.

\subsubsection{Regions III and IV}

Here the number of independent parameters diminishes: $SU(2)_R$ invariance
implies $m_d = m_u \equiv m_{Q^c}$, $m_e = m_N \equiv m_{L^c}$, $m_{H_u} =
m_{H_d} \equiv m_\Phi$, $h_u = h_d \equiv h_q$ and $h_e = h_N \equiv h_l$ at
energies $\geq M_R$:
\begin{eqnarray} \label{soft2_A}
\beta_{h_q} &=& h_q \left[ {\rm tr}(3Y_qY_q^\dagger + Y_l Y_l^\dagger) +
    5 Y_q Y_q^\dagger + Y_q^\dagger Y_q - \frac{16}{3}g_3^2 - 3g_2^2 -
      3g_R^2 - \frac{g_{B-L}^2}{6} \right] 
      \nonumber\\ 
&+& Y_q \left[ {\rm tr}(6 h_q Y_q^\dagger + 2h_l Y_l^\dagger) + 
 4 h_q Y_q^\dagger + 2 Y_q^\dagger h_q 
+ \frac{32}{3}g_3^2M_3 + 6g_2^2M_2 + 6g_R^2M_R +
      \frac{g_{B-L}^2}{3}M_{B-L} \right] \,; \nonumber \\
\beta_{h_l} &=& h_l \left[ {\rm tr}(3 Y_q Y_q^{\dagger} + Y_l Y_l^{\dagger}) +
      5 Y_l Y_l^{\dagger} + Y_l^{\dagger} Y_l + \frac{3}{2}Y_N^{\dagger}Y_N -
      3g_2^2 - 3g_R^2 - \frac{3}{2}g_{B-L}^2 \right] \nonumber\\
 &+& Y_l \left[ {\rm tr}(6 h_q Y_q^{\dagger} + 2 h_l Y_l^{\dagger}) +
 4 h_l Y_l^{\dagger} + 2 Y_l^{\dagger} h_l + 3Y_N^{\dagger}h_N 
 + 6g_2^2M_2 + 6g_R^2M_R + 3g_{B-L}^2 \right] \,; \nonumber \\
\beta_{h_N} &=& h_N \left[ \frac{1}{2}{\rm tr}(Y_N^{\dagger}Y_N) + 4Y_l
Y_l^{\dagger} + 3Y_N Y_N^{\dagger} - 7g_R^2 - \frac{9}{2}g_{B-L}^2 \right]
      \nonumber\\  
&+& Y_N \left[ Y_N^{\dagger}h_N + 8Y_l^{\dagger} h_l + 6Y_N^{\dagger}h_N +
      14g_R^2M_R + 9g_{B-L}^2 \right]  \,.
\end{eqnarray}

\begin{eqnarray} \label{soft2_B}
\beta_{B} &=& B \left[{\rm tr}(6 Y_q Y_q^{\dagger} + 2 Y_l Y_l^{\dagger}) 
- 3g_2^2 - 3g_R^2 \right]
+ \mu \left[ {\rm tr}(12 h_q Y_q^\dagger + 4 h_l Y_l^\dagger ) 
+ 6g_2^2M_2 + 6g_R^2M_R \right] \,; \nonumber \\
\beta_{B_\Sigma} &=& B_\Sigma \left[ \frac{1}{2}{\rm tr}(Y_N Y_N^\dagger) 
- 8g_R^2 - 6g_{B-L}^2 \right] + M_\Sigma \left[ {\rm tr}(Y_N^{\dagger}h_N)
 + 16g_R^2M_R + 12g_{B-L}^2M_{B-L} \right] \,. \nonumber \\
\end{eqnarray}

\begin{eqnarray} \label{soft2_m}
\beta_{m^2_\Phi} &=& {\rm tr} \left[ 6(m^2_\Phi + m^2_Q)Y_q^{\dagger}Y_q +
      6Y_q^{\dagger}m^2_{Q^c} Y_q + 2(m^2_\Phi + m^2_L)Y_l^{\dagger}Y_l +
     2Y_l^{\dagger}m^2_{L^c} Y_l \right. \nonumber\\
&& \left. \hspace*{5mm} + 6h_q^{\dagger}h_q + 2h_l^{\dagger}h_l \right]
- 6g_2^2|M_2|^2 - 6g_R^2|M_R|^2 \,; \nonumber \\
\beta_{m^2_{\bar{\Sigma}}} &=&
\frac{1}{2}{\rm tr}(Y_N^{\dagger}Y_Nm^2_{\bar{\Sigma}} +
 Y_NY_N^{\dagger}m^2_{\bar{\Sigma}}) + {\rm tr}(2Y_N^{\dagger}m_{L^c}^2Y_N +
  h_N^{\dagger}h_N) \nonumber\\
&-& 16|M_R|^2g_R^2 - 12|M_{B-L}|^2g_{B-L}^2 + 3g_{B-L}^2 S_{B-L}\,; 
\nonumber \\
\beta_{m^2_{\Sigma}} &=& - 16|M_R|^2g_R^2 - 12|M_{B-L}|^2g_{B-L}^2
      - 3g_{B-L}^2 S_{B-L}\,; \nonumber \\
\beta_{m^2_Q} &=& 2 (m^2_Q + 2m^2_\Phi)Y_q^{\dagger}Y_q 
      + 2 Y_q^{\dagger}Y_q m^2_Q
 + 4Y_q^{\dagger}m^2_{Q^c} Y_q  \nonumber\\
&+& 4h_q^{\dagger} h_q - \frac{32}{3}g_3^2|M_3|^2 - 
6g_2^2|M_2|^2 - \frac{1}{3}g_{B-L}^2|M_{B-L}|^2 + \frac{1}{2}g_{B-L}^2S_{B-L} 
\,; \nonumber \\
\beta_{m^2_L} &=& 2 (m^2_L + 2m^2_\Phi)Y_l^{\dagger}Y_l 
      + 2 Y_l^{\dagger} Y_l m^2_L
+ 4 Y_l^{\dagger}m^2_l Y_l \\
&+&      4h_l^{\dagger}h_l 
- 6g_2^2|M_2|^2 - 3g_{B-L}^2|M_{B-L}|^2 - \frac{3}{2}g_{B-L}^2S_{B-L}
\,; \nonumber \\
\beta_{m^2_{Q^c}} &=& (2m^2_{Q^c} + 4m^2_\Phi) Y_q Y_q^{\dagger} +
    4Y_q m^2_Q Y_q^{\dagger} + 2Y_q Y_q^{\dagger}m_{Q^c}^2 + 4h_qh_q^{\dagger}
   \nonumber\\ 
&-& \frac{32}{3}g_3^2|M_3|^2 - 6g_R^2|M_R|^2 -
\frac{1}{3}g_{B-L}^2|M_{B-L}|^2 - \frac{1}{2}g_{B-L}^2S_{B-L} \,; \nonumber \\
\beta_{m^2_{L^c}} &=& (2m_{L_c}^2 + 4m^2_\Phi)Y_l Y_l^{\dagger} +
      4Y_l m^2_L Y_l^{\dagger} + 2Y_l Y_l^{\dagger} m_{L^c}^2 
 + 4h_l h_l^{\dagger} + \frac{3}{2} m_{L^c}^2 Y_NY_N^{\dagger} 
+ \frac{3}{2}Y_NY_N^{\dagger}m_{L^c}^2 
\nonumber\\ 
 &+& 3 Y_N^{\dagger} m_{L^c}^2 Y_N + 
      3Y_N^{\dagger} m^2_{\bar{\Sigma}} Y_N + 3h_N^{\dagger}h_N
- 6g_R^2|M_R|^2 - 3g_{B-L}^2|M_{B-L}|^2 + \frac{3}{2}g_{B-L}^2S_{B-L}
 \,. \nonumber
\end{eqnarray}
Note that $\Sigma$, which we introduced to allow a $D-$flat direction for
symmetry breaking at scale $M_R$, does not have any superpotential couplings,
hence its soft mass does not appear in any of the other RGE. We again list its
RGE for completeness.

\subsubsection{Region V}

At scales above $M_C$ the spectrum further simplifies: $G_{422}$ invariance
implies that $m_Q = m_L \equiv m_F$, $m_{Q^c} = m_{L^c} = m_{F^c}$, and $h_q =
h_l \equiv h$. In addition, the discrete $D$ symmetry implies
$m_{\overline{\Sigma}_L} = m_{\overline \Sigma_R} \equiv m_{\overline
  \Sigma}$ and $m_F = m_{F^c}$:
\begin{eqnarray} \label{soft3_A}
\beta_h &=& h \left[4 {\rm tr} ( Y Y^{\dagger}) +
\frac{15}{4} Y_N Y_N^{\dagger} +  5 Y Y^{\dagger} + Y^{\dagger}Y -
\frac{15}{2}g_4^2 - 3g_2^2 -   3g_R^2 \right] 
 \nonumber\\ 
&+&  Y \left[ {\rm tr}(8 h Y^{\dagger} ) +
  \frac{15}{2}h_N Y_N^{\dagger} + 4 h Y^{\dagger} + 2 Y^{\dagger} h
+ 15g_4^2M_4 + 6g_2^2M_2 + 6g_R^2M_R \right] \,;\nonumber\\
\beta_{h_N} &=& h_N \left[\frac{1}{2}{\rm tr}(Y_N^{\dagger}Y_N) + 4 Y
      Y^{\dagger} + \frac{15}{2} Y_N Y_N^{\dagger} - \frac{33}{2}g_4^2 -
      7g_R^2 \right] \nonumber\\  
&+& Y_N \left[{\rm tr}(h_N Y_N^\dagger) + 8 Y^\dagger h 
+ 15 Y_N^\dagger h_N + 33 g_4^2M_4 + 14 g_R^2M_R \right]  \,.
\end{eqnarray}

\begin{eqnarray} \label{soft3_B}
\beta_{B} &=& B \left[{\rm tr}(8 Y Y^{\dagger}) - 3g_2^2 - 3g_R^2 \right]
+ \mu \left[{\rm tr}(16 h Y^\dagger) 
      + 6g_2^2M_2 + 6g_R^2M_R \right] \,;\nonumber\\
\beta_{B_\Sigma} &=& B_\Sigma \left[\frac{1}{2}{\rm tr}(Y_N Y_N^\dagger) -
      8g_R^2 - 18g_4^2 \right] + m_\Sigma \left[{\rm tr}(Y_N^{\dagger}h_N)
   + 16g_R^2M_R + 36g_4^2M_4 \right] \,.
\end{eqnarray}

\begin{eqnarray} \label{soft3_m}
\beta_{m^2_{\Phi}} &=& {\rm tr} \left[8 (m^2_{\Phi} + m^2_F) Y^{\dagger} Y +
      8 Y^{\dagger} m^2_F Y 
+ 8h_u^\dagger h_u \right] 
- 6g_2^2|M_2|^2 - 6g_R^2|M_R|^2 \,;\nonumber\\
\beta_{m^2_{\overline \Sigma}} &=&
      \frac{1}{2}{\rm tr}(Y_N^{\dagger} Y_N m^2_{\overline \Sigma} +
      Y_N Y_N^{\dagger} m^2_{\overline \Sigma}) +{\rm tr}(2 Y_N^{\dagger}
      m_F^2 Y_N + h_N^{\dagger}h_N)
   - 16|M_R|^2g_R^2 - 36|M_4|^2 g_4^2 \,;\nonumber\\
\beta_{m^2_{\Sigma}} &=&   - 16|M_R|^2g_R^2 - 36|M_4|^2g_4^2 \,;\nonumber
\\
\beta_{m^2_F} &=& 2 (m^2_F + 2m^2_{\Phi}) Y^{\dagger} Y 
      + 2 Y^{\dagger} Y m^2_F 
+ \frac{15}{4}(m_F^2 + 2m^2_{\bar{\Sigma}}) Y_N Y_N^{\dagger} +
      \frac{15}{4} Y_N Y_N^{\dagger} m_F^2
\nonumber\\
  &+& \frac{15}{2}Y_N 
      m_F^2 Y_N^{\dagger} + \frac{15}{2}h_Nh_N^{\dagger}
+  4 Y^{\dagger} m^2F Y + 4 h^{\dagger} h 
       - 15g_4^2|M_4|^2 - 6g_2^2|M_2|^2 \,.
    \end{eqnarray}

\end{document}